\documentclass[12pt]{article}

\usepackage{latexsym,amsmath,amssymb,amsthm,amsfonts,graphicx,natbib}
\usepackage{epsfig}
\usepackage{setspace}
\usepackage{bm}
\usepackage{hyperref}
\usepackage[autolinebreaks,useliterate]{mcode}

\newcommand*\patchAmsMathEnvironmentForLineno[1]{%
      \expandafter\let\csname old#1\expandafter\endcsname\csname #1\endcsname
      \expandafter\let\csname oldend#1\expandafter\endcsname\csname end#1\endcsname
      \renewenvironment{#1}%
         {\linenomath\csname old#1\endcsname}%
         {\csname oldend#1\endcsname\endlinenomath}}%
    \newcommand*\patchBothAmsMathEnvironmentsForLineno[1]{%
      \patchAmsMathEnvironmentForLineno{#1}%
      \patchAmsMathEnvironmentForLineno{#1*}}%
    \AtBeginDocument{%
    \patchBothAmsMathEnvironmentsForLineno{equation}%
    \patchBothAmsMathEnvironmentsForLineno{align}%
    \patchBothAmsMathEnvironmentsForLineno{flalign}%
    \patchBothAmsMathEnvironmentsForLineno{alignat}%
    \patchBothAmsMathEnvironmentsForLineno{gather}%
    \patchBothAmsMathEnvironmentsForLineno{multline}%
    }
\usepackage[mathlines,displaymath]{lineno}	
\usepackage[textwidth=7in,textheight=9in,centering]{geometry}

\include{def}
\def\dispmuskip{\thinmuskip= 3mu plus 0mu minus 2mu \medmuskip=  4mu plus 2mu minus 2mu \thickmuskip=5mu plus 5mu minus 2mu}
\def\textmuskip{\thinmuskip= 0mu                    \medmuskip=  1mu plus 1mu minus 1mu \thickmuskip=2mu plus 3mu minus 1mu}
\def\beq{\dispmuskip\begin{equation}}    \def\eeq{\end{equation}\textmuskip}
\def\beqn{\dispmuskip\begin{displaymath}}\def\eeqn{\end{displaymath}\textmuskip}
\def\bea{\dispmuskip\begin{eqnarray}}    \def\eea{\end{eqnarray}\textmuskip}
\def\bean{\dispmuskip\begin{eqnarray*}}  \def\eean{\end{eqnarray*}\textmuskip}

\newtheorem{algorithm}{Algorithm}

\newcommand{\diag}{\text{diag}}

\newcommand{\eps}{\epsilon}
\newcommand{\wh}{\widehat}
\newcommand{\wt}{\widetilde}
\def\trace{\text{\rm tr}}

\def\E{{\mathbb E}}                         % Expectation
\def\V{{\mathbb V}}
\def\a{\alpha}

\def\d{\rm d}
\def\eps{\epsilon}
\def\veps{\varepsilon}

\def\s{\sigma}

\def\t{\theta}
\def\b{\beta}
\def\l{\lambda}

\def\F{{\cal F}}

\def\N{{\cal N}}

\def\KL{\text{\rm KL}}

\def\vec{\text{\rm vec}}
\def\vech{\text{\rm vech}}

\def\cov{\text{\rm cov}}

\def\LB{\text{\rm LB}}

\def\diag{\text{\rm diag}}
\def\Var{\text{\rm Var}}

\newcommand{\bol}[1]{\textbf{#1}}

\theoremstyle{definition} % to make the text within examples non-italic. 
\newtheorem{examplex}{Example}[section]
\newenvironment{example}
  {\pushQED{\qed}\examplex} % add a triangle symbol at the end of example.
  {\popQED\endexamplex}

\begin{document}
%\doublespacing
\title{A practical tutorial on Variational Bayes}
\author{Minh-Ngoc Tran, Trong-Nghia Nguyen, Viet-Hung Dao
\footnote{Tran and Nguyen are with the University of Sydney Business School. Dao is with the UNSW Business School. Correspondence to minh-ngoc.tran@sydney.edu.au.
The authors would like to thank David Nott and Emtiyaz Khan for useful comments and suggestions. We also thank Robert Salomone for pointing out many errors in an early version. Any errors left are our own.}}
\date{\empty}
\maketitle
\begin{abstract}
This tutorial gives a quick introduction to Variational Bayes (VB), also called Variational Inference or Variational Approximation, from a practical point of view. 
The paper covers a range of commonly used VB methods and an attempt is made 
to keep the materials accessible to the wide community of data analysis practitioners.
The aim is that the reader can quickly derive and implement their first VB algorithm for Bayesian inference with their data analysis problem. 
An end-user software package in Matlab together with the documentation can be found at \url{https://vbayeslab.github.io/VBLabDocs/}
\end{abstract}

%===========================================================%
\section{Introduction}\label{sec:Introduction}
%===========================================================%
Bayesian inference has been long called for Bayesian computation techniques that are scalable to large data sets 
and applicable in big and complex models with a huge number of unknown parameters to infer.  
Sampling methods, such as Markov Chain Monte Carlo (MCMC) and Sequential Monte Carlo (SMC),  in their current development do not meet this need.
Sampling methods have not been successfully used in some modern areas such as deep neural networks.
Even in more traditional areas such as graphical modelling and mixture modelling, it is very challenging to use MCMC and SMC.  
Variational Bayes (VB) is an optimization-based technique for approximate Bayesian inference, and provides a computationally efficient alternative to sampling methods.
VB belongs to the bigger class of Variational Inference methods, which can
also be used in the frequentist context for maximum likelihood estimation
when there are missing data. The names Variational Bayes and Variational
Inference are often used exchangeably in the literature, however, we prefer
the former in this tutorial as we are solely interested in approximating the
posterior distributions for Bayesian inference.

This tutorial provides a quick introduction to VB.
There are many excellent tutorials and review papers on VB,
however, most of them are either too abstract or tangential to the statistics readership,
and do not offer much hands-on experience.
This tutorial focuses on the practical aspect of VB,
and is written to help the reader, who might even have a little background in computational statistics, 
be able to quickly learn about VB and implement the method to fit their model.

Let $y$ denote the data and $p(y|\theta)$ the likelihood function based on a postulated model, with $\theta\in\Theta$ the vector of model parameters to be estimated.
Let $p(\theta)$ be the prior. Bayesian inference encodes all the available information about the model parameter $\theta$ in its posterior distribution with density
\[p(\theta|y)=\frac{p(y,\theta)}{p(y)}=\frac{p(\theta)p(y|\theta)}{p(y)}\propto p(\theta)p(y|\theta),\]
where $p(y)=\int_\Theta p(\theta)p(y|\theta)d\theta$, called the {\it marginal likelihood} or {\it evidence}.
Here, the notation `$\propto$' means proportional up to the normalizing constant that is independent of the parameter ($\theta$).
In most Bayesian derivations, such a constant can be safely ignored.
Bayesian inference typically requires computing expectations with respect to the posterior distribution.
For example, the posterior mean, which is often used for point estimation, is an expectation of $\theta$ with respect to the posterior distribution $p(\theta|y)$.  
However, it is often difficult to compute such expectations, partly because the density $p(\theta|y)$ itself is intractable as the normalizing constant $p(y)$ is often unknown.
For many applications, Bayesian inference is performed using MCMC, which estimates 
expectations w.r.t. $p(\theta|y)$ by sampling from it.
For other applications where $\theta$ is high dimensional or fast computation is of primary interest, VB is an attractive alternative to MCMC. 
VB approximates the posterior distribution by a probability distribution with density $q(\theta)$
belonging to some tractable family of distributions $\mathcal Q$ such as Gaussians. The best VB approximation $q^*\in\mathcal Q$ is found by minimizing the Kullback-Leibler (KL) divergence {\it from} $q(\theta)$ {\it to} $p(\theta|y)$ 
\beq\label{eq: original VB prob}
q^*=\arg\min_{q\in\mathcal Q}\left\{ \KL\big(q\|p(\cdot|y)\big):=\int q(\theta)\log\frac{q(\theta)}{p(\theta|y)}d\theta\right\}.
\eeq
Then, Bayesian inference is performed with the intractable posterior $p(\theta|y)$ replaced by the tractable VB approximation $q^*(\theta)$.
It is easy to see that
\[\KL(q\|p(\cdot|y)) = -\int q(\theta)\log\frac{p(\theta)p(y|\theta)}{q(\theta)}d\theta+\log p(y),\]
thus minimizing  KL is equivalent to maximizing the lower bound on $\log p(y)$\footnote{In this tutorial, the notation $a:=b$ means $a$ is defined by $b$. 
For any random variable or random vector $X$ and any function $g(X)$, we denote by $\E_{f}\big(g(X)\big)$ (or $\E_{X\sim f}\big(g(X)\big)$, or simply $\E_{X}\big(g(X)\big)$) the expectation of $g(X)$ where $X$ follows a probability distribution with density function $f(x)$.}
\begin{equation}\label{eq: original LB}
\LB(q):=\int q(\theta)\log\frac{p(\theta)p(y|\theta)}{q(\theta)}d\theta=\E_{q}\Big(\log\frac{p(\theta)p(y|\theta)}{q(\theta)}\Big).
\end{equation}

Without any constraint on $\mathcal Q$, the solution to \eqref{eq: original VB prob} is $q^*(\theta)=p(\theta|y)$; of course this solution is useless as it is itself intractable.
Depending on the constraint imposed on the class $\mathcal Q$, VB algorithms can be categorized into two classes: Mean Field VB (MFVB) and Fixed Form VB (FFVB)
which are presented in Section \ref{sec:MFVB} and Section \ref{sec:FFVB}, respectively. These two sections can be read completely separately depending on the reader's interest.

For researchers who wish to reproduce the numerical results in this tutorials,
the Matlab code together with the data used in the examples are available on our gibhub \url{https://github.com/VBayesLab/Tutorial-on-VB}.
For general practitioners, we provide an end-user software package VBLab, also available on our gibhub site, that allows users to easily 
perform approximate Bayesian inference in a wide range of statistical models. 
Section \ref{sec:software package} describes this user-friendly VBLab software package and its applications.

%===========================================================%
\section{Mean Field Variational Bayes}\label{sec:MFVB}
%===========================================================%
Let's write $\t$ as $\t=(\t_1^\top,\t_2^\top)^\top$. Here $a^\top$ denotes the transpose of vector $a$; and all vectors in this tutorial are column vectors.
MFVB assumes the following factorization form for $q$
\[q(\t)=q_1(\t_1)q_2(\t_2),\]
i.e., we ignore the posterior dependence between $\theta_1$, $\theta_2$ and attempt to approximate $p(\t_1,\t_2|y)$ by $q(\t)=q_1(\t_1)q_2(\t_2)$. This is the only assumption/restriction we put on the class $\mathcal Q$.
The lower bound in \eqref{eq: original LB} is 
\bean
\LB(q_1,q_2)&=&\int q_1(\t_1)q_2(\t_2)\log\frac{p(\t,y)}{q_1(\t_1)q_2(\t_2)}d\t_1d\t_2\\
&=&\int q_1(\t_1)q_2(\t_2)\log p(\t,y)d\t_1d\t_2\\
&&-\int q_1(\t_1)\log q_1(\t_1)d\t_1-\int q_2(\t_2)\log q_2(\t_2)d\t_2\\
&=&\int q_1(\t_1)\E_{-\t_1}[\log p(y,\t)]d\t_1-\int q_1(\t_1)\log q_1(\t_1)d\t_1+C(q_2)
\eean
where $\E_{-\t_1}[\log p(y,\t)]:=\E_{q_2(\t_2)}[\log p(y,\t)]=\int q_2(\t_2)\log p(y,\t) d\t_2$ and $C(q_2)$ is the term independent of $q_1$.
The funny-looking notation $\E_{-\t_1}(\cdot)$,
meaning we take the expectation with respect to everything except $\theta_1$, turns out to be very convenient when we deal with the general MFVB procedure later.
Hence,
\bea
\LB(q_1,q_2)&=&\int q_1(\t_1)\log\frac{\exp\big(\E_{-\t_1}[\log p(y,\t)]\big)}{q_1(\t_1)}d\t_1+C(q_2)\notag\\
&=&\int q_1(\t_1)\log\frac{\wt q_1(\theta_1)}{q_1(\t_1)}d\t_1+C(q_2)+\log\wt C(q_2)\notag\\
&=&-\KL(q_1\|\wt q_1)+C(q_2)+\log\wt C(q_2),
\eea
where $\wt q_1(\t_1)$ is the probability density function determined by
\[\wt q_1(\t_1):=\frac{\exp(\E_{-\t_1}[\log p(y,\t)])}{\wt C(q_2)}\propto \exp(\E_{-\t_1}[\log p(y,\t)]),\]
with $\wt C(q_2):=\int \exp(\E_{-\t_1}[\log p(y,\t)])d\t_1$ also independent of $q_1$.
We therefore have that
\beq\label{eq: MFVB 1}
\LB(q_1,q_2)=-\KL(q_1\|\wt q_1)+\text{ constant independent of $q_1$}.
\eeq
Similarly,
\beq\label{eq: MFVB 2}
\LB(q_1,q_2)=-\KL(q_2\|\wt q_2)+\text{ constant independent of $q_2$},
\eeq
where $\wt q_2(\t_2)\propto \exp(\E_{-\t_2}[\log p(y,\t)])$ with $\E_{-\t_2}[\log p(y,\t)]:=\int q_1(\t_1)\log p(y,\t) d\t_1$.
The expressions in \eqref{eq: MFVB 1}-\eqref{eq: MFVB 2} suggest a coordinate ascent optimization procedure for maximizing the lower bound: given $q_2$, we minimize $\KL(q_1\|\wt q_1)$ to find $q_1$, and given $q_1$ we minimize $\KL(q_2\|\wt q_2)$ to find $q_2$.
The hope is that solving the optimization problems
\beq\label{eq: MFVB coordinate ascent}
\min_{q_1}\big\{\KL(q_1\|\wt q_1)\big\}\;\;\;\text{ and }\;\;\;\min_{q_2}\big\{\KL(q_2\|\wt q_2)\big\}
\eeq
is easier than minimizing the original KL divergence between $q(\theta_1,\theta_2)$ and $p(\theta_1,\theta_2|y)$.
If $\wt q_1$ and $\wt q_2$ are tractable and standard distributions\footnote{By a standard distribution, or a recognizable distribution, we mean a probability distribution that is well-understood and widely used, such as Gaussian, Gamma, etc. Yes, this definition of standard distribution isn't standard!}, then of course the solution to \eqref{eq: MFVB coordinate ascent} is 
$q_1=\wt q_1$ and $q_2=\wt q_2$.
The most useful scenario is the case of {\it conjugate prior}: the prior $p(\t_1)$ belongs to a parametric density family $\F_1$, then
$\wt q_1(\t_1)$ also belongs to $\F_1$. Similarly,  the prior $p(\t_2)$ belongs to a parametric density family $\F_2$, then
$\wt q_2(\t_2)$ also belongs to $\F_2$.
Then the solutions to \eqref{eq: MFVB coordinate ascent} are
\[q_1(\theta_1)=\wt q_1(\t_1)\in\F_1\;\;\;\text{ and }\;\;\;q_2(\theta_2)=\wt q_2(\t_2)\in\F_2,\]
and in order to identify $q_1$ and $q_2$ it's only necessary to compute their parameters.
Computing the parameter in $q_1$ requires $q_2$ and vice versa,
which suggests the following coordinate ascent-type algorithm for maximizing the lower bound:     
\begin{algorithm}[Mean Field Variational Bayes]\label{al: algorithm 1}
	
\begin{itemize}
\item[1.] Initialize the parameter of $q_1(\t_1)$

\item[2.] Given $q_1(\t_1)$, update the parameter of $q_2(\t_2)$ using
\beq\label{eq:optimal VB 1}
q_2(\t_2)\propto \exp\big(\E_{-\t_2}[\log p(y,\t)]\big)=\exp\Big(\int q_1(\t_1)\log p(y,\t_1,\t_2)d\t_1\Big).
\eeq
\item[3.] Given $q_2(\t_2)$, update the parameter of $q_1(\t_1)$ using 
\beq\label{eq:optimal VB 2}
q_1(\t_1)\propto \exp\big(\E_{-\t_1}[\log p(y,\t)]\big)=\exp\Big(\int q_2(\t_2)\log p(y,\t_1,\t_2)d\t_2\Big).
\eeq
\item[4.] Repeat Steps 2 and 3 until the stopping condition is met.
\end{itemize}

\end{algorithm}
A stopping rule is to terminate the update if the change in the parameters of the VB posterior $q(\t)=q_1(\t_1)q_2(\t_2)$ between two consecutive iterations is less than some threshold $\epsilon$.
In the case the lower bound $\LB(q_1,q_2)$ can be computed, one can stop the algorithm if the increase (or the percentage of the increase) in the lower bound is less than some threshold.
Note that $\LB(q)$ increases after each iteration.

\begin{example}\label{exa:Example 1}
Let $ y=(11; 12; 8; 10; 9; 8; 9; 10; 13; 7)$ be observations from $\N(\mu,\s^2)$, the normal distribution with mean $\mu$ and variance $\s^2$. Suppose that we use the prior $\N(\mu_0,\s_0^2)$ for $\mu$ and $\text{Inverse-Gamma}(\a_0,\b_0)$ for $\s^2$,
with hyperparameters $\mu_0=0$, $\s_0=10$, $\a_0=1$ and $\b_0=1$. 
Assume the VB factorization $q(\mu,\sigma^2)=q(\mu)q(\sigma^2)$.
Let's derive the MFVB procedure for  
approximating the posterior $p(\mu,\s^2| y)\propto p(\mu)p(\s^2)p( y|\mu,\s^2)$. We can view $\mu$ and $\sigma^2$ respectively as $\theta_1$ and $\theta_2$ in Algorithm \ref{al: algorithm 1}.

From \eqref{eq:optimal VB 1}, the optimal VB posterior for $\s^2$ is
\bean
q(\s^2)&\propto&\exp\Big(\E_{-\s^2}[\log p( y,\mu,\s^2)]\Big)=\exp\Big(\E_{q(\mu)}[\log p( y,\mu,\s^2)]\Big)\\
&\propto&\exp\Big(\E_{q(\mu)}[\log p(\s^2)+\log p( y|\mu,\s^2)]\Big)\\
&\propto&\exp\Big(-(\a_0+\frac n2+1)\log\s^2-\big(\b_0+\frac12\E_{q(\mu)}[\sum(y_i-\mu)^2]\big)/\s^2\Big).
\eean
In the above derivation, we have ignored all the constants independent of $\sigma^2$ as they are unnecessary for identifying the distribution $q(\sigma^2)$.
It follows that $q(\s^2)$ is inverse-Gamma with parameters
\[\a_q=\a_0+\frac n2,\;\;\;\;\b_q=\b_0+\frac12\E_{q(\mu)}\Big[\sum(y_i-\mu)^2\Big].\] 
Computation of the expecation $\E_{q(\mu)}(\cdot)$ becomes clear shortly after $q(\mu)$ is identified.
From \eqref{eq:optimal VB 2}, the optimal VB posterior for $\mu$ is
\bean
q(\mu)&\propto&\exp\Big(\E_{q(\s^2)}[\log p(y,\mu,\s^2)]\Big)\\
&\propto&\exp\Big(\E_{q(\s^2)}[\log p(\mu)+\log p( y|\mu,\s^2)]\Big)\\
&\propto&\exp\Big(-\frac{1}{2\s_0^2}(\mu^2-2\mu_0\mu)-\frac n2\E_{q(\s^2)}[\frac{1}{\s^2}](-2\bar y\mu+\mu^2)\Big)\\
&\propto&\exp\Big(-\frac{1}{2}\underbrace{\big(\frac{1}{\s_0^2}+n\E_{q(\s^2)}[\frac{1}{\s^2}]\big)}_{A}\mu^2+\mu\underbrace{\big(\frac{\mu_0}{\s_0^2}+n\bar y\E_{q(\s^2)}[\frac{1}{\s^2}]\big)}_B\Big)\\
&=&\exp\Big(-\frac{1}{2}A\mu^2+B\mu\Big)\\
&\propto&\exp\Big(-\frac12\frac{(\mu-{B}/{A})^2}{1/A}\Big).
\eean
It follows that $q(\mu)$ is Gaussian with mean $\mu_q$ and variance $\s_q^2$
\[\mu_q=\frac{\frac{\mu_0}{\s_0^2}+n\bar y\E_{q(\s^2)}[\frac{1}{\s^2}]}{\frac{1}{\s_0^2}+n\E_{q(\s^2)}[\frac{1}{\s^2}]},\;\;\;\;\;
\s_q^2=\Big(\frac{1}{\s_0^2}+n\E_{q(\s^2)}[\frac{1}{\s^2}]\Big)^{-1}.\]
With the distributions $q(\mu)$ and $q(\sigma^2)$ having identified, we are now able to compute the expectations w.r.t. $q(\mu)$ and $q(\sigma^2)$ in the above:
\bean
\b_q&=&\b_0+\frac12\E_{q(\mu)}\big[\sum(y_i-\mu)^2\big]\\
&=&\b_0+\frac12\Big(\sum y_i^2-2n\bar y\E_{q(\mu)}[\mu]+n\E_{q(\mu)}[\mu^2]\Big)\\
&=&\b_0+\frac12\sum y_i^2-n\bar y\mu_q+\frac{n}{2}(\mu_q^2+\s_q^2).
\eean
As $q(\s^2)\sim\text{Inverse-Gamma}(\a_q,\b_q)$, $\E(1/\s^2)=\a_q/\b_q$. Hence,
\[\mu_q=\Big(\frac{\mu_0}{\s_0^2}+n\bar y\frac{\a_q}{\b_q}\Big)/\Big(\frac{1}{\s_0^2}+n\frac{\a_q}{\b_q}\Big),\;\;\text{ and }\;\;\s_q^2=\Big(\frac{1}{\s_0^2}+n\frac{\a_q}{\b_q}\Big)^{-1}.\]
Note that we did not make any assumption on the parametric form of optimal variational distributions $q(\mu)$ and $q(\sigma^2)$,
it is the model (the prior and the likelihood) that determines their form. 
We arrive at the following updating procedure:
\begin{itemize}
\item Initialize $\mu_q,\s_q^2$
\item Update the following recursively 
\bean
\a_q&\leftarrow&\a_0+\frac n2,\\
\b_q&\leftarrow&\b_0+\frac12\sum y_i^2-n\bar y\mu_q+\frac{n}{2}(\mu_q^2+\s_q^2),\\
\mu_q&\leftarrow&\Big(\frac{\mu_0}{\s_0^2}+n\bar y\frac{\a_q}{\b_q}\Big)/\Big(\frac{1}{\s_0^2}+n\frac{\a_q}{\b_q}\Big),\\
\s_q^2&\leftarrow&\Big(\frac{1}{\s_0^2}+n\frac{\a_q}{\b_q}\Big)^{-1},
\eean
until convergence.
\end{itemize}
We can stop the iterative scheme when the change of the $\ell_2$-norm of the vector $\l=(\a_q,\b_q,\mu_q,\s_q^2)^\top$ is smaller than some $\eps$, $\eps=10^{-5}$ for example.
We can also initialize $\a_q,\b_q$ and then update the variational parameters recursively in the order of $\mu_q$, $\s_q^2$, $\a_q$ and $\b_q$. However, it's often a better idea to initialize $\mu_q,\s_q^2$ as it is easier to guess the values related to location parameters than the scale parameters.
Figure \ref{fig:MFVB} plots the posterior densities estimated by the MFVB algorithm derived above, and by Gibbs sampling.  

\begin{figure}[h]
\centering
\includegraphics[width=1\columnwidth]{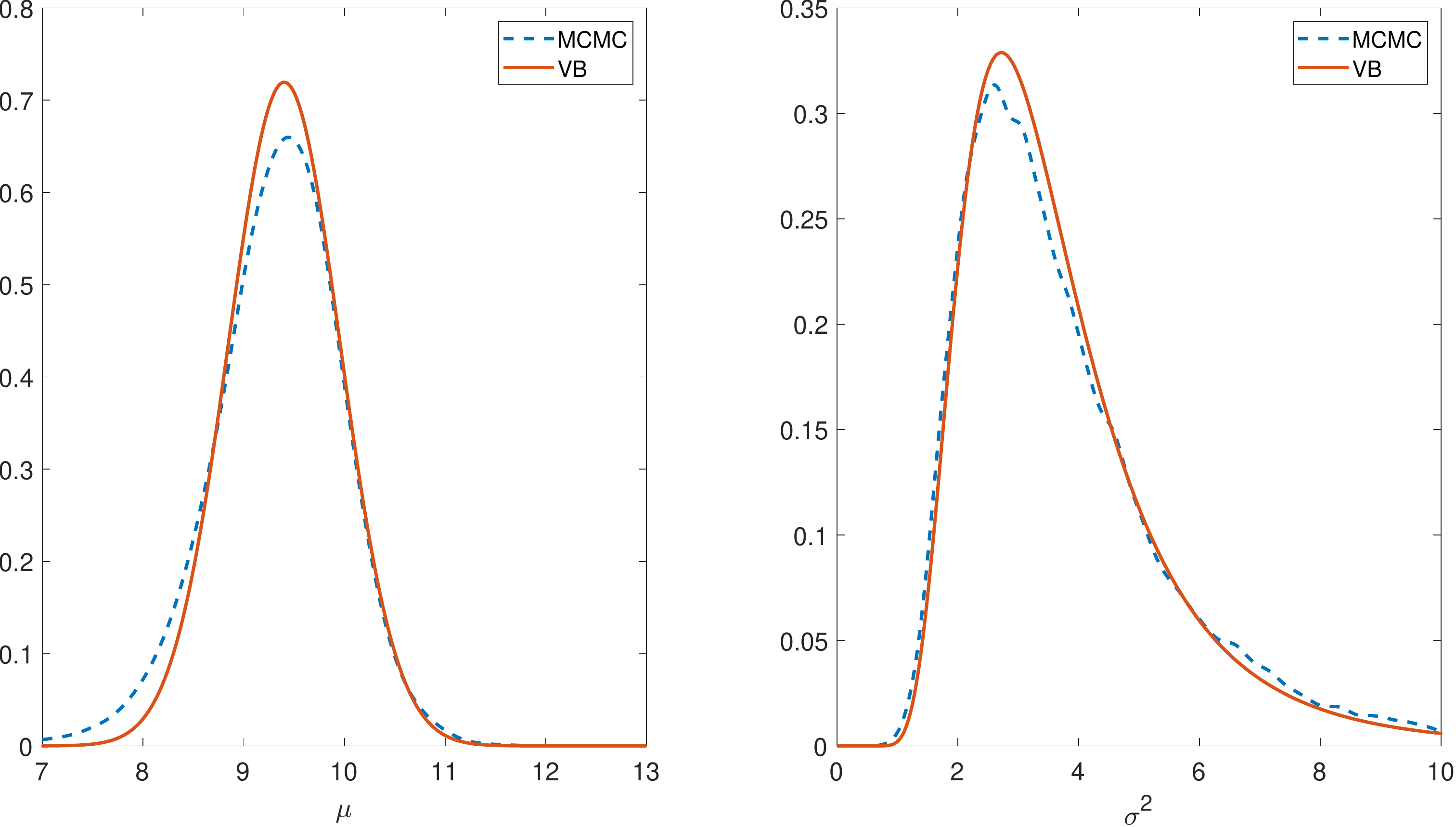}
\caption{Example \ref{exa:Example 1}: Posterior density for $\mu$ and $\sigma^2$ estimated by MFVB and Gibbs sampling. The CPU time
taken by VB was 0.006 seconds, by the Gibbs sampling scheme was 1.81 seconds. VB was about 300
times faster.} 
\label{fig:MFVB}
\end{figure}
\end{example}

It is straightforward to extend the MFVB procedure in Algorithm \ref{al: algorithm 1} to the general case where $\theta$ is divided into $k$ blocks $\t=(\t_1^\top,\t_2^\top,...,\t_k^\top)^\top$,
and where we want to approximate the posterior $p(\t_1,\t_2,...,\t_k|y)$ by $q(\t)=q_1(\t_1)q_2(\t_2)...q_k(\t_k)$.
The optimal $q_j(\t_j)$ that maximizes $\LB(q)$, when $q_1,...,q_{j-1},q_{j+1},...,q_{k}$ are fixed, is
\beq\label{eq:optimal VB}
q_j(\t_j)\propto \exp\big(\E_{-\t_j}[\log p(y,\t)]\big),\;\;\;j=1,...,k.
\eeq
Here $\E_{-\t_j}(\cdot)$ denotes the expectation w.r.t. $q_1$,..., $q_{j-1}$, $q_{j+1}$,..., $q_{k}$, i.e.,
\[\E_{-\t_j}\big[\log p(y,\t)\big]:=\int q_1(\theta_1)...q_{j-1}(\theta_{j-1})q_{j+1}(\theta_{j+1})...q_{k}(\theta_k)\log p(y,\t)d\theta_1....d\theta_{j-1}d\theta_{j+1}...d\theta_k.\]
A similar procedure to Algorithm \ref{al: algorithm 1} can be developed, in which we first initialize the parameters in the $k-1$ factors $q_1,...,q_{k-1}$, then update $q_k$ and the other factors recursively.

%------------------------------------------------------------------------%
\subsection{MFVB for elaborate models}\label{sec:ChapterMFVB:Elaborate Distributions}
%------------------------------------------------------------------------%
One of the difficulties in using MFVB is that the optimal variational distributions in \eqref{eq:optimal VB} sometimes do not admit a standard form.
In Example \ref{exa:Example 1}, for example, if the data $y_i$ does not follow a normal distribution but a Student's $t$ distribution $t_\nu(\mu,\sigma^2)$,
then it can be seen that the optimal variational distribution $q(\mu)$ does not have the form of a Gaussian distribution or any standard probability distribution.
In some situations, however, by introducing auxiliary variables, we can equivalently represent the model by augmenting the parameter space  
such that MFVB is applicable.
The use of auxiliary variables to facilitate statistical computations is widely used in many areas of statistics.
We follow \cite{wand2011} and use the term {\it elaborate model} to refer to a statistical model 
in which its prior or its data density can be augmented using auxiliary variables such that the optimal variational distributions 
in \eqref{eq:optimal VB} admit a standard form.
Introducing auxiliary variables makes MFVB tractable, but this might come at the price of reducing the variational approximation accuracy; however, we won't discuss this issue in any detail in this tutorial.

More precisely, consider the standard Bayesian model
\beq
y|\theta\sim p(y|\theta),\;\;\quad\quad
\theta\sim p(\theta)\label{eq:ChaperMFVB:model 1}.
\eeq
Suppose that there exists an auxiliary variable $\eta$ such that
\beq
p(y|\theta) = \int p(y|\theta,\eta)p(\eta|\theta)\d\eta,
\eeq
then model \eqref{eq:ChaperMFVB:model 1} can be equivalently represented as
\beq
y|\theta,\eta\sim p(y|\theta,\eta),\;\;\quad\quad
\eta|\theta\sim p(\eta|\theta),\;\;\quad\quad
\theta\sim p(\theta)\label{eq:ChaperMFVB:model 2}.
\eeq
The model \eqref{eq:ChaperMFVB:model 1} is said to be elaborate if it can be presented as 
the hierarchical model \eqref{eq:ChaperMFVB:model 2} and, under the variational factorization $q(\theta,\eta)=q(\theta)q(\eta)$, the optimal variational distributions $q(\theta)$ and $q(\eta)$ in \eqref{eq:optimal VB} admit a standard form. The idea of elaborate models applies to the prior too, in which one can represent the prior $p(\theta)$ in a hierarchical form using auxiliary variables.

We now demonstrate this idea in the Bayesian Lasso model.
Consider the linear regression problem
\beqn
y =\mu 1_n +X\beta+\epsilon,
\eeqn 
where $y$ is the vector of responses, $X$ is the $n \times p$ matrix of covariates,
$1_n$ is the $n \times 1$ vector of 1s, and $\epsilon$ is the vector of i.i.d. normal errors $\N(0,\sigma^2)$.
Without loss of generality, we assume that $y$ and $X$ have been centered so that $\mu$ is zero and omitted from the model.
Regression analysis is often concerned with estimating $\beta=(\beta_1,...,\beta_p)^\top$ and simultaneously identifying the important covariates.
The least absolute shrinkage and selection operator (Lasso) method solves this problem by minimizing the sum of squared errors and a regularization term
\beq\label{eq:ChapterMFVB:lasso}
\underset{\beta}{\mbox{min}} ~\Big\{(y-X\beta)'(y-X\beta)+\wt\lambda \sum_{j=1}^p |\beta_j|\Big\},
\eeq
where $\wt\lambda >0$ is the tuning parameter controlling the amount of regularization. 
The Lasso estimator, i.e. the solution of \eqref{eq:ChapterMFVB:lasso}, can be interpreted as the posterior mode in a Bayesian context
where a conditional Laplace prior is used for $\beta$
\beq\label{eq:ChapterMFVB:prior}
p(\beta|\sigma^2) = \prod_{j=1}^p \frac{\lambda}{2\sqrt{\sigma^2}} e^{-\lambda|\beta_j|/\sqrt{\sigma^2}},
\eeq
for some shrinkage parameter\footnote{This parameter shouldn't be confused with the variational parameter $\lambda$ in Section \ref{sec:FFVB}.} $\lambda$.
The posterior mode of $\beta$ is the Lasso estimator in \eqref{eq:ChapterMFVB:lasso} with $\wt\lambda=2\sqrt{\sigma^2}\lambda$.

It is difficult to use MFVB for approximating the posterior $p(\beta,\sigma^2|X,y)$ in this case,
as the optimal conditional variational distribution of $\beta$ does not admit a standard form.
However, it turns out that we can use auxiliary variables to make this Bayesian model elaborate and overcome the aforementioned difficulty.

It is well-known that a Laplace distribution can be represented as a mixture of normal and exponential distributions as follows
\beqn
\frac{\lambda}{2} e^{-\lambda|z|}
=\int_0^\infty \frac{1}{\sqrt{2\pi s}}e^{-z^2/(2s)}\frac{\lambda^2}{2}
e^{-\lambda^2s/2}\d s.
\eeqn
Using this representation, after some algebra, we have that
\beqn
\frac{\lambda}{2\sqrt{\sigma^2}} e^{-\lambda|\beta_j|/\sqrt{\sigma^2}}=\int_0^\infty \frac{1}{\sqrt{2\pi\sigma^2\tau}}e^{-\beta_j^2/({2\sigma^2\tau})}\frac{\lambda^2}{2}
e^{-\lambda^2\tau/2}\d \tau.
\eeqn
This motivates the following hierarchical representation of the Bayesian Lasso model
\bean\label{eq2}
y|X,\beta,\sigma^2 &\sim & \N(X\beta,\sigma^2 I_n),\\
\beta_j|\sigma^2,\tau_j&\sim&\N(0,\sigma^2\tau_j),\\
\tau_j&\sim& \text{Exp}\big(\frac{\lambda^2}{2}\big) =\frac{\lambda^2}{2}e^{-\lambda^2\tau_j/2},\;\;\;j=1,...,p.\label{eq:ChapterMFVB:tau prior}
\eean
The conjugate prior for $\sigma^2$ is inverse Gamma and we use the improper prior $p(\sigma^2)\propto 1/\sigma^2$ in this example.
The shrinkage parameter $\lambda$ can be selected in some way, here we use a full Bayesian treatement and put a Gamma prior on $\lambda^2$
\beqn\label{eq:ChapterMFVB:prior lambda}
p(\lambda^2)=\frac{\delta^r}{\Gamma(r)}(\lambda^2)^{r-1}e^{-\delta\lambda^2},
\eeqn
with $r$ and $\delta$ hyperparameters and pre-specified.
Note that we use a prior for $\lambda^2$, not $\lambda$, as this leads to a tractable form for the optimal conditional variational distribution for $\lambda^2$.

The model parameters include $\beta$, $\tau=(\tau_1,...,\tau_p)^\top$, $\sigma^2$ and $\lambda^2$.
Let us use the following mean field variational distribution
\beqn
q(\beta,\tau,\sigma^2,\lambda^2)=q(\beta)q(\tau)q(\sigma^2)q(\lambda^2).
\eeqn
With this factorization, all the optimal conditional variational distributions admit a standard form.
The optimal variational distribution for $\beta$ is $\N(\mu_\beta,\Sigma_\beta)$ with
\beqn
\mu_\beta=\big(X^\top X+D_\tau\big)^{-1}X^\top y,\;\;\;\;\Sigma_\beta=\big(X^\top X+D_\tau\big)^{-1}/\E_q(\frac{1}{\sigma^2}),
\eeqn
where $D_\tau:=\diag\big(\E_q(1/\tau_1),\cdots,\E_q(1/\tau_p)\big)$. Here $\E_q(\cdot)$ denotes expectation with respect to the variational distribution $q$.
The optimal variational distributions for $\tau_j$ are independent of each other,
where $\wt\tau_j:= 1/\tau_j$ follows an inverse-Gaussian with location and scale parameters
\beqn
\mu_{\wt\tau_j} = \Big(\frac{\E_q(\lambda^2)}{\E_q\big(\beta_j^2/\sigma^2\big)}\Big)^{1/2},\;\;\;\;\lambda_{\wt\tau_j} = \E_q(\lambda^2).
\eeqn
The optimal distribution for $\sigma^2$ is inverse Gamma with the parameters
\beqn
\alpha_{\sigma^2}=\frac12(n+p),\;\;\;\;\beta_{\sigma^2}=\frac12\E_q\|y-X\beta\|^2+\frac12\sum_{j=1}^p\E_q\big(\frac{\beta_j^2}{\tau_j}\big).
\eeqn
Finally, the optimal variatinoal distribution for $\lambda^2$ is Gamma with
\beqn
\alpha_{\lambda^2}= r+1,\;\;\;\; \beta_{\lambda^2} = \delta+\frac12\sum_j\E_q(\tau_j).
\eeqn
Using the results regarding to the moments of these standard distributions, we have
\begin{align*}
\E_q\big(\frac{1}{\tau_j}\big)&=\E_q(\wt\tau_j)=\mu_{\wt\tau_j},&\E_q(\tau_j)&=\E_q\big(\frac{1}{\wt\tau_j}\big)=\frac{1}{\mu_{\wt\tau_j}}+\frac{1}{\lambda_{\wt\tau_j}},\\
\E_q\big(\frac{1}{\sigma^2}\big)&=\frac{\alpha_{\sigma^2}}{\beta_{\sigma^2}},&\E_q\big(\lambda^2\big)&=\frac{\alpha_{\lambda^2}}{\beta_{\lambda^2}},\\
\E_q(\beta_j^2)&=\mu_{\beta,j}^2+\Sigma_{\beta,jj},
\end{align*}
where $\mu_{\beta,j}$ is the $j$th element of vector $\mu_{\beta}$ and $\Sigma_{\beta,jj}$ is the $(j,j)$ element of matrix $\Sigma_{\beta}$.
We arrive at the MFVB procedure for Bayesian inference in the Bayesian Lasso model.

\begin{algorithm}[MFVB for Bayesian Lasso]\label{al:ChapterMFVB:algorithm 2}Initialize $\alpha_{\sigma^2}$, $\beta_{\sigma^2}$, $\mu_{\wt\tau_j}$ and $\lambda_{\wt\tau_j}$, $j=1,...,p$, then update the following until convergence:
\begin{itemize}
\item Update $\mu_\beta$ and $\Sigma_\beta$
\beqn
\mu_\beta=\big(X^\top X+D_\tau\big)^{-1}X^\top y,\;\;\;\;\Sigma_\beta=\frac{\beta_{\sigma^2}}{\alpha_{\sigma^2}}\big(X^\top X+D_\tau\big)^{-1},
\eeqn
where $D_\tau:=\diag\big(\mu_{\wt\tau_1},\cdots,\mu_{\wt\tau_p}\big)$. 
\item Update $\alpha_{\lambda^2}$ and $\beta_{\lambda^2}$ 
\beqn
\alpha_{\lambda^2}= r+1,\;\;\;\; \beta_{\lambda^2} = \delta+\frac12\sum_j\Big(\frac{1}{\mu_{\wt\tau_j}}+\frac{1}{\lambda_{\wt\tau_j}}\Big).
\eeqn
\item Update $\mu_{\wt\tau_j}$ and $\lambda_{\wt\tau_j}$, $j=1,...,p$
\beqn
\mu_{\wt\tau_j} = \Big(\frac{\alpha_{\lambda^2}/\beta_{\lambda^2}}{\big(\alpha_{\sigma^2}/\beta_{\sigma^2}\big)\big(\mu_{\beta,j}^2+\Sigma_{\beta,jj}\big)}\Big)^{1/2},\;\;\;\;\lambda_{\wt\tau_j} = \frac{\alpha_{\lambda^2}}{\beta_{\lambda^2}}.
\eeqn
\item Update $\alpha_{\sigma^2}$ and $\beta_{\sigma^2}$
\beqn
\alpha_{\sigma^2}=\frac12(n+p),\;\;\;\;\beta_{\sigma^2}=\frac12\|y-X\mu_{\beta}\|^2+\frac12\trace(X\Sigma_\beta X^\top)+\frac12\sum_{j=1}^p(\mu_{\beta,j}^2+\Sigma_{\beta,jj})\mu_{\wt\tau_j}.
\eeqn
\end{itemize}
\end{algorithm}

%------------------------------------------------------------------------%
\begin{example}[Bayesian Lasso]\label{exa:Bayesian Lasso}
A data set of size $n=500$ is generated from the model
\beqn
y=x^\top\beta+\sigma\epsilon,
\eeqn
where $\beta=(3,\ 1.5,\ 0,\ 0,\ 2,\ 0,\ 0,\ 0)^\top$, $x_j\sim \N(0,1)$,
$\epsilon\stackrel{iid}{\sim}\N(0,1)$ and $\sigma=0.1$.

The MFVB algorithm \ref{al:ChapterMFVB:algorithm 2} stops after 22 iterations
when the $l_2$ difference between two consecutive updates of $\mu_\beta$ is less than $1e-10$.
The hyperparameters $r$ and $\delta$ are set to 0.
Table \ref{tab:ChapterMFVB:BLasso} summarizes the result,
Figure \ref{fig:ChapterMFVB:BLasso1} plots the updates of $\mu_\beta$ over iterations. 

\begin{table}
  \begin{center}
    \begin{tabular}{cc}
    \hline\hline
True $\beta$&$\mu_\beta$\\
\hline
3&3.0029 (0.0042)\\
   1.5&    1.4946 (0.0041)\\
         0&    0.0044 (0.0040)\\
         0   & 0.0074 (0.0043)\\
    2  &  2.0064 (0.0041)\\
         0  & $-0.0088$ (0.0044)\\
         0 &  $-0.0007$ (0.0041)\\
         0  &  0.0008 (0.0039)\\
\hline\hline    
    \end{tabular}
  \end{center}
  \caption{Example \ref{exa:Bayesian Lasso}: The performance of MFVB for the Bayesian Lasso model. The first column lists the true $\beta$ and the second column lists the point estimate $\mu_\beta$ (at convergence) of the posterior mean of $\beta$, with the estimates of the posterior standard deviations in brackets.} \label{tab:ChapterMFVB:BLasso}
\end{table}

\begin{figure}[h]
\centering
\includegraphics[width=1\textwidth,height=.4\textheight]{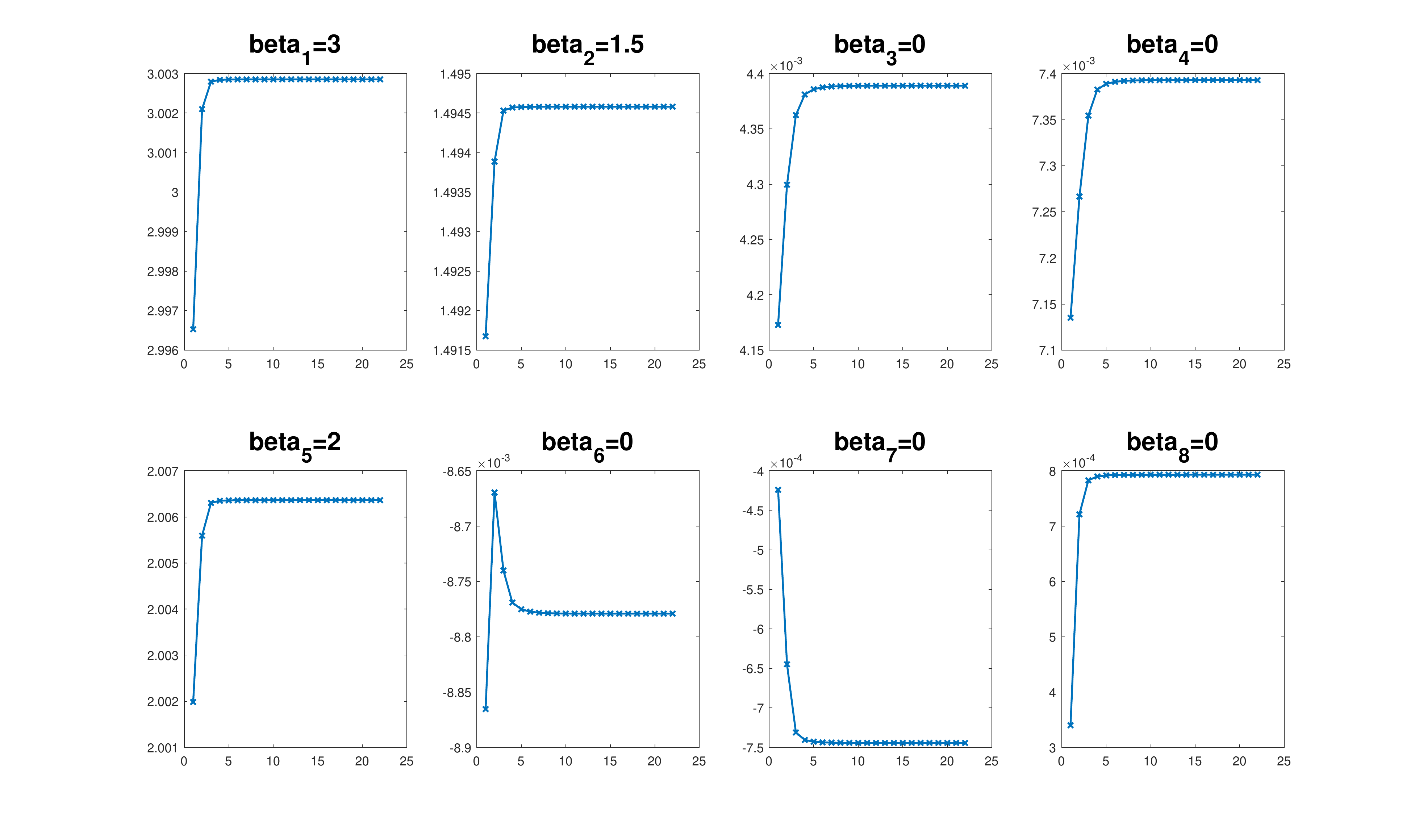}
\caption{Example \ref{exa:Bayesian Lasso}: The updates of $\mu_\beta$ over iterations.} 
\label{fig:ChapterMFVB:BLasso1}
\end{figure}

\end{example}

%------------------------------------------------------------------------%
\subsection{Some remarks about MFVB}\label{sec:MFVB:remarks}
%------------------------------------------------------------------------%
Early work on MFVB in machine learning and statistics can be found in \cite{Waterhouse:1996,Jordan.et.all:1999} and \cite{Titterington:2004},
and tutorial-style introductions to MFVB can be found in \cite{Bishop:2006} and \cite{Ormerod:2010}.
MFVB has been successfully used in some statistical areas such as mixture modelling and graphical modelling.
In mixture modelling, for example, MFVB does not only offer a fast Bayesian estimation method, but is also able to deal with the challenging model selection problem in a convenient way.
See, e.g., \cite{Ghahramani:2000,Corduneanu:2001,McGrory:2007,Giordani.Mun.Tran.Kohn:2013} and \cite{Tran.Giordani.Kohn.Pitt:2014}.
The approximation accuracy and large-scale properties of MFVB have been extensively studied recently, its cover requires a book-length discussion and is omitted in this tutorial.

%===========================================================%
\section{Fixed Form Variational Bayes}\label{sec:FFVB}
%===========================================================%
FFVB assumes a fixed parametric form for the VB approximation density $q$, i.e. $q=q_\l$ belongs to some class of distributions $\mathcal Q$ indexed by a vector $\lambda$ called the {\it variational parameter}.
For example, $q_\l$ is a Gaussian distribution with mean $\mu$ and covariance matrix $\Sigma$.
FFVB finds the best $q_\lambda$ in the class $\mathcal{Q}$ by optimizing the lower bound
\beq\label{eq: LB}
\LB(\lambda):= \LB(q_\lambda) =\E_{q_\lambda}\left[\log\frac{p(\theta)p(y|\theta)}{q_\lambda(\theta)}\right]=\E_{q_\lambda}\big[h_\lambda(\theta)\big],
\eeq
with 
\beqn
h_\lambda(\t):=\log\big(\frac{p(\t) p(y|\t)}{q_\lambda(\theta)}\big).
\eeqn
Later we also use $h(\theta)$, without the subscript, to denote the model-specific function $\log\;\big(p(\t) p(y|\t)\big)$.
Except for a few trivial cases where the LB can be computed analytically and optimized using classical optimization routines,
stochastic optimization is often used to optimize $\LB(\lambda)$.
The gradient vector of LB is
\bea
\nabla_\lambda\LB(\lambda)&=&\int_\Theta\nabla_\lambda q_\lambda(\theta)\log\frac{p(\theta)p(y|\theta)}{q_\lambda(\theta)}d\theta-\int_\Theta q_\lambda(\theta)\nabla_\lambda\log q_\lambda(\theta)d\theta\notag\\
&=&\int_\Theta q_\lambda(\theta)\nabla_\lambda \log q_\lambda(\theta)\log\frac{p(\theta)p(y|\theta)}{q_\lambda(\theta)}d\theta-\int_\Theta \nabla_\lambda q_\lambda(\theta)d\theta\notag\\
&=&\int_\Theta q_\lambda(\theta)\nabla_\lambda \log q_\lambda(\theta)\log\frac{p(\theta)p(y|\theta)}{q_\lambda(\theta)}d\theta-\nabla_\lambda\int_\Theta  q_\lambda(\theta)d\theta\notag\\
&=&\E_{q_\lambda}\left[\nabla_\lambda \log q_\lambda(\theta)\times\log\frac{p(\theta)p(y|\theta)}{q_\lambda(\theta)}\right]\notag\\
&=&\E_{q_\lambda}\left[\nabla_\lambda \log q_\lambda(\theta)\times h_\lambda(\theta)\right]\label{eq:lb gradient}.
\eea
The gradient in this form is often referred to as {\it score-function gradient},
another way known as {\it reparameterization gradient} to compute the gradient of the lower bound is discussed later in \eqref{eq:Reparameterization trick gradient}.
It follows from \eqref{eq:lb gradient} that, by generating\footnote{In Monte Carlo simulation, by $\t\sim q_\lambda(\t)$ we mean that we draw a random variable or random vector $\theta$ from the probability distribution with density $q_\l(\t)$. That notation also means $\theta$ is a random variable/vector whose probability density function is $q_\l(\t)$.} $\t\sim q_\lambda(\t)$,  it is straightforward to obtain an unbiased estimator $\wh{\nabla_\l\text{LB}}(\l)$ of the gradient $\nabla_\l\text{LB}(\l)$, i.e., $\E\big[\wh{\nabla_\l\text{LB}}(\l)\big]=\nabla_\l\text{LB}(\l)$.
Therefore, we can use stochastic optimization\footnote{Unbiased estimate of the gradient of the target function is theoretically required in stochastic optimization.} to optimize $\text{LB}(\lambda)$.
The basic algorithm is as follows:
\begin{algorithm}[Basic FFVB algorithm]\label{algorithm 1}

\begin{itemize}
  \item Initialize $\l^{(0)}$ and stop the following iteration if the stopping criterion is met.
  \item For $t=0,1,...$
  \begin{itemize}
	  \item Generate $\theta_s\sim q_{\lambda^{(t)}}(\theta)$, $s=1,...,S$
	  \item Compute the unbiased estimate of the LB gradient
	  \[\wh{\nabla_\l\text{LB}}(\l^{(t)}):=\frac{1}{S}\sum_{s=1}^S\nabla_\lambda \log q_\lambda(\theta_s)\times h_\lambda(\theta_s)|_{\lambda=\lambda^{(t)}}.\]
	  \item Update 
	  \bea\label{eq:update rule}
	  \l^{(t+1)}=\l^{(t)}+a_t\wh{\nabla_\l\text{LB}}(\l^{(t)}).
	  \eea
  \end{itemize}
\end{itemize}
\end{algorithm}
The algorithmic parameter $S$ is referred to as the number of Monte Carlo samples (used to estimate the gradient of the lower bound).
The sequence of learning rates $\{a_t\}$ should satisfy the theoretical requirements $a_t>0$, $\sum_t a_t=\infty$ and $\sum_t a_t^2<\infty$.
However, this basic VB algorithm hardly works in practice and requires some refinements to make it work.
Much of the rest of this section focuses on presenting and explaining those refinements.

%--------------------------------------------------------------------%
\subsection{Stopping criterion} 
%--------------------------------------------------------------------%
Let us first discuss on the stopping rule. An easy-to-implement stopping rule is to terminate the updating
procedure if the change between $\lambda^{(t+1)}$ and $\lambda^{(t)}$, e.g. in terms of the Euclidean distance, is less than some threshold $\epsilon$.
However, it is difficult to select a meaningful $\epsilon$ as such a distance depends on the scales and the length of the vector $\l$.
Denote by $\wh{\text{LB}}(\l)$ an estimate of $\text{LB}(\l)$ by sampling from $q_\lambda(\theta)$, i.e.,
\beqn
\wh{\text{LB}}(\l)=\frac{1}{S}\sum_{s=1}^S h_{\lambda}(\theta_s),\quad \theta_s\sim q_\lambda(\theta).
\eeqn
Although $\text{LB}(\l)$ is expected to be non-decreasing over iterations,
its sample estimate $\wh{\text{LB}}(\l)$ might not be.
To account for this, we can use a moving average of the lower bounds
over a window of $t_W$ iterations, $\overline {\text{LB}}(\l^{(t)})=(1/t_W)\sum_{k=1}^{t_W} \wh{\text{LB}}(\l^{(t-k+1)})$.
At convergence, the values $\text{LB}(\l^{(t)})$ stay roughly the same,
therefore $\overline {\text{LB}}(\l^{(t)})$ will average out the noise in $\wh{\text{LB}}(\l^{(t)})$ and is stable. 
The stopping rule that is widely used in machine learning is to stop training if the moving averaged lower bound does not improve
after $P$ iterations; and $P$ is sometimes fancily referred to as the {\it patience} parameter. 
Typical choice is $P=20$ or $P=50$, and $t_W=20$ or $t_W=50$. Note that, we must not use the last $\lambda^{(t)}$ as the final estimate of $\lambda$, but the one corresponding to the largest $\overline {\text{LB}}(\l^{(t)})$.

\subsection{Adaptive learning rate and natural gradient}
Let's write the update in \eqref{eq:update rule} as
\[\begin{cases}
	  \l^{(t+1)}_1=\l^{(t)}_1+a_t\wh{\nabla_{\l_1}\text{LB}}(\l^{(t)})\\
	  ...\\
  	  \l^{(t+1)}_{d_\l}=\l^{(t)}_{d_\l}+a_t\wh{\nabla_{\l_{d_\l}}\text{LB}}(\l^{(t)}),	  
\end{cases}\]
with $d_\l$ the size of vector $\lambda$, which shows that a common scalar learning rate $a_t$ is used for all the coordinates of $\lambda$.
Intuitively, each coordinate of vector $\lambda^{(t+1)}$
might need a different learning rate that can take into account the scale of that coordinate or the geometry of the space $\lambda$ living in.
It turns out that the basic Algorithm \ref{algorithm 1} rarely works in practice without a method for selecting the learning rate adaptively.

\subsubsection{Adaptive learning rate}\label{sec: Adaptive learning rate}
For a coordinate $i$ with a large variance $\V(\wh{\nabla_{\l_i}\text{LB}}(\l^{(t)}))$, its learning rate $a_{t,i}$ should be small, otherwise the new update $\lambda^{(t+1)}_i$ jumps all over the place and destroys everything the process has learned so far.
Denote $g_t:=\wh{\nabla_{\lambda}\LB}(\l^{(t)})$ be the gradient vector at step $t$,
and $v_t:=(g_t)^2$ (this is a coordinate-wise operator).
The commonly used adaptive learning rate methods such as ADAM and AdaGrad work by scaling the coordinates of $g_t$ by their corresponding variances.
These variances are estimated by moving average.
The algorithm below is a basic version of this class of adaptive learning methods: 
\begin{itemize}
\item[1)] Initialize $\l^{(0)}$, $g_0$ and $v_0$ and set $\bar g=g_0$, $\bar v=v_0$. Let $\beta_1,\beta_2\in(0,1)$ be adaptive learning weights.
\item[2)] For $t=0,1,...$, update
\bean
\bar g &=&\beta_1 \bar g+(1-\beta_1)g_t\\
\bar v &=&\beta_2 \bar v+(1-\beta_2)v_t\\
\l^{(t+1)}&=&\l^{(t)}+\a_t \bar g/\sqrt{\bar v},
\eean
with $\a_t$ a scalar step size. Here $\bar g/\sqrt{\bar v}$ should be understood component wise.
\end{itemize}
Note that the LB gradients $g_t$ have also been smoothened out using moving average.
This helps to accelerate the convergence - a method known as the momentum method in the stochastic optimization literature.
Typical choice of the scalar $\a_t$ is
\beq\label{eq:scalar learning rate}
\a_t=\min\left(\eps_0,\eps_0\frac{\tau}{t}\right)=\begin{cases}
\eps_0,&t\leq\tau\\
\eps_0\frac{\tau}{t},&t>\tau
\end{cases}
\eeq
for some small {\it fixed learning rate} $\eps_0$ (e.g. 0.1 or 0.01) and some threshold $\tau$ (e.g., 1000).  
In the first $\tau$ iterations, the training procedure explores the learning space with a fixed learning rate $\eps_0$,
then this exploration is settled down by reducing the step size after $\tau$ iterations.
%\subsubsection*{ADADELTA:}
%\subsubsection*{ADAM:}
%\begin{example}\label{ex: }	
%\end{example}
\subsubsection{Natural gradient}
Natural gradient can be considered as an adaptive learning method that exploits the geometry of the $\lambda$ space.
The ordinary gradient $\nabla_\l{\LB}(\l)$ does not adequately capture the geometry of the approximating family $\mathcal Q$ of $q_\l(\t)$.
A small Euclidean distance between $\l$ and $\l'$ does not necessarily mean a small \KL{} divergence between $q_\l(\t)$ and $q_{\l'}(\t)$.
Statisticians and machine learning researchers have long realized the importance of information geometry on the manifold of a statistical model,
and that the steepest direction for optimizing the objective function $\LB (\l)$ on the manifold formed by the family $q_\l(\t)$ is directed by the so-called natural gradient 
which is defined by pre-multiplying the ordinary gradient with the inverse of the Fisher information matrix
\beqn
\nabla_{\lambda}\LB (\l)^{\text{nat}} := I_F^{-1}(\l)\nabla_\l\LB(\l),
\eeqn
with  $I_F(\lambda)=\cov_{q_\l}(\nabla_\l\log q_\l(\t))$ the Fisher information matrix about $\lambda$ with respect to the distribution $q_\lambda$. Given an unbiased estimate $\wh{\nabla_\l\text{LB}}(\l)$, the unbiased estimate of the natural gradient is 
\beq\label{eq:natural gradient}
\wh{\nabla_{\lambda}\LB}(\l)^{\text{nat}} = I_F^{-1}(\l)\wh{\nabla_\l\LB}(\l).
\eeq

The main difficulty in using the natural gradient is the computation of $I_F(\lambda)$, and the solution of 
the linear systems required to compute \eqref{eq:natural gradient}. 
The problem is more severe in high dimensional models because this matrix has a large size.
An efficient method for computing $I_F(\lambda)^{-1}\wh{\nabla_\l \LB}(\l)$ is using iterative conjugate gradient methods which solve the linear system $I_F(\lambda)x=\wh{\nabla_\lambda \LB} (\l)$ for $x$ using only matrix-vector products involving $I_F(\lambda)$.
In some cases this matrix vector product can be done efficiently both in terms of computational time and memory requirements by exploiting the structure
of the Fisher matrix $I_F(\lambda)$. 
See Section \ref{sec: GVB factor} for a special case where the natural gradient is computed efficiently in high dimensional problems.

As mentioned before, the gradient momentum method is often useful in stochastic optimization that 
helps accelerate and stabilize the optimization procedure.
The momentum update rule with the natural gradient is 
\bean
\overline{{\nabla_\l{\LB}}} &=& \alpha_\text{m} \overline{{\nabla_\l{\LB}}}+(1-\alpha_\text{m})\wh{\nabla_{\lambda}\LB}(\l^{(t)})^{\text{nat}},\\
\l^{(t+1)}&=&\l^{(t)}+\a_t \overline{{\nabla_\l{\LB}}},
\eean
where $\alpha_\text{m}\in[0,1]$ is the momentum weight; $\alpha_m$ around 0.6-0.9 is a typical choice. 
The use of the moving average gradient $\overline{{\nabla_\l{\LB}}}$ also helps remove some of the noise
inherent in the estimated gradients of the lower bound.
Note that the momentum method is already embedded in the moving-average-based adaptive learning rate methods in Section \ref{sec: Adaptive learning rate}. 

%-------------------------------------------------------------%
\subsection{Control variate}\label{sec:control variate}
%-------------------------------------------------------------%
As is typical of stochastic optimization algorithms,
the performance of Algorithm \ref{algorithm 1} depends greatly on the variance of the noisy gradient.
Variance reduction for the noisy gradient is a key ingredient in FFVB algorithms.   
This section describes a control variate technique for variance reduction,
another technique known as {\it reparameterization trick} is presented in Section \ref{reparam trick}.

Let $\t_s\sim q_\l(\t)$, $s=1,...,S$, be $S$ samples from the variational distribution $q_{\l}(\t)$. A naive estimator of the $i$th element of the vector $\nabla_\lambda\text{LB}(\l)$ is
\beq\label{eq: naive KL estimate}
\wh{\nabla_{\lambda_i}\LB}(\l)^{\text{naive}}=\frac1S\sum_{s=1}^S\nabla_{\lambda_i}[\log q_\lambda(\t_s)]\times h_\lambda(\t_s),
\eeq
whose variance is often too large to be useful. 
For any number $c_i$, consider   
\beq\label{eq: reduced var KL estimate}
\wh{\nabla_{\lambda_i}\LB}(\l)=\frac1S\sum_{s=1}^S\nabla_{\lambda_i}[\log q_\lambda(\t_s)](h_\lambda(\t_s)-c_i),
\eeq
which is still an unbiased estimator of $\nabla_{\lambda_i}\text{LB}(\l)$ since $\E(\nabla_{\lambda}[\log q_\lambda(\t)])=0$,
whose variance can be greatly reduced by an appropriate choice of control variate $c_i$.
The variance of $\wh{\nabla_{\lambda_i}\LB}(\l)$ is 
\beqn
\frac1S\V\Big(\nabla_{\lambda_i}[\log q_\lambda(\t)]h_\lambda(\t)\Big)+\frac{c_i^2}{S}\V\Big(\nabla_{\lambda_i}[\log q_\lambda(\t)]\Big)-\frac{2c_i}{S}\cov\Big(\nabla_{\lambda_i}[\log q_\lambda(\t)]h_\lambda(\t),\nabla_{\lambda_i}[\log q_\lambda(\t)]\Big).
\eeqn
The optimal $c_i$ that minimizes this variance is 
\beq\label{eq:optimal c_i}
c_i=\cov\Big(\nabla_{\lambda_i}[\log q_\lambda(\t)] h_\lambda(\t),\nabla_{\lambda_i}[\log q_\lambda(\t)]\Big)\Big/\V\Big(\nabla_{\lambda_i}[\log q_\lambda(\t)]\Big).
\eeq
Then $\V(\wh{\nabla_{\lambda_i}\LB}(\l))=\V(\wh{\nabla_{\lambda_i}\LB}(\l)^{\text{naive}})(1-\rho^2_i)\leq\V(\wh{\nabla_{\lambda_i}\LB}(\l)^{\text{naive}})$,
where $\rho_i$ is the correlation between $\nabla_{\lambda_i}[\log q_\lambda(\t)]h_\lambda(\t)$ and $\nabla_{\lambda_i}[\log q_\lambda(\t)]$.
Often, $\rho_i^2$ is very close to 1, which leads to a large variance reduction.

One can estimate the numbers $c_i$ in \eqref{eq:optimal c_i} using samples $\t_s\sim q_{\l}(\t)$.
In order to ensure the unbiasedness of the gradient estimator, the samples used to estimate $c_i$ must be independent of the samples used to estimate the gradient.
In practice, the $c_i$ can be updated sequentially as follows.
At iteration $t$, we use the $c_i$ computed in the previous iteration $t-1$, i.e. based on the samples from $q_{\l^{(t-1)}}(\t)$,
to estimate the gradient $\wh{\nabla_\l\text{LB}}(\l^{(t)})$,
which is computed using new samples from $q_{\l^{(t)}}(\t)$.
We then update the $c_i$ using this new set of samples.
By doing so, the unbiasedness is guaranteed while no extra samples are needed in updating the control variates $c_i$.

Algorithm \ref{algorithm 2} provides a detailed pseudo-code implementation of the FFVB approach that uses the control variate for variance reduction 
and moving average adaptive learning,
and Algorithm \ref{algorithm 3} implements the FFVB approach that uses the control variate and natural gradient.  

\begin{algorithm}[FFVB with control variates and adaptive learning]\label{algorithm 2} 

{\bf Input}: Initial $\l^{(0)}$, adaptive learning weights $\beta_1,\beta_2\in(0,1)$, fixed learning rate $\eps_0$, threshold $\tau$, rolling window size $t_W$ and maximum patience $P$. {\bf Model-specific requirement}: function $h(\theta):=\log\big(p(\theta)p(y|\theta)\big)$.
\begin{itemize}
  \item Initialization
  \begin{itemize}
	  \item Generate $\theta_s\sim q_{\lambda^{(0)}}(\theta)$, $s=1,...,S$.
	  \item Compute the unbiased estimate of the LB gradient
	  \[\wh{\nabla_\l\text{LB}}(\l^{(0)}):=\frac{1}{S}\sum_{s=1}^S\nabla_\lambda \log q_\lambda(\theta_s)\times h_\lambda(\theta_s)|_{\lambda=\lambda^{(0)}}.\]
	  \item Set $g_0:=\wh{\nabla_\l\text{LB}}(\l^{(0)})$, $v_0:=(g_0)^2$, $\bar g:=g_0$, $\bar v:=v_0$. 
	  \item Estimate the vector of control variates $c$ as in \eqref{eq:optimal c_i} using the samples $\{\theta_s,s=1,...,S\}$.
	  \item Set $t=0$, $\text{patience}=0$ and \texttt{stop=false}.
  \end{itemize}
  \item While \texttt{stop=false}:
  \begin{itemize}
	  \item Generate $\theta_s\sim q_{\lambda^{(t)}}(\theta)$, $s=1,...,S$.
	  \item Compute the unbiased estimate of the LB gradient\footnote{The term $\nabla_\lambda \log q_\lambda(\theta_s)\circ \big(h_\lambda(\theta_s)-c\big)$ should be understood component-wise, i.e. it is the vector whose $i$th element is $\nabla_{\lambda_i} \log q_\lambda(\theta_s)\times \big(h_\lambda(\theta_s)-c_i\big)$.}
	  \[g_t:=\wh{\nabla_\l\text{LB}}(\l^{(t)})=\frac{1}{S}\sum_{s=1}^S\nabla_\lambda \log q_\lambda(\theta_s)\circ \big(h_\lambda(\theta_s)-c\big)|_{\lambda=\lambda^{(t)}}.\]
	  \item Estimate the new control variate vector $c$ as in \eqref{eq:optimal c_i} using the samples $\{\theta_s,s=1,...,S\}$.
	  \item Compute $v_t=(g_t)^2$ and 
	  \[\bar g =\beta_1 \bar g+(1-\beta_1)g_t,\;\;\bar v =\beta_2 \bar v+(1-\beta_2)v_t.\]
	  \item Compute $\alpha_t=\min(\epsilon_0,\epsilon_0\frac{\tau}{t})$ and update
	  \[\l^{(t+1)}=\l^{(t)}+\a_t \bar g/\sqrt{\bar v}\]
	  \item Compute the lower bound estimate
	  \[\wh{\text{LB}}(\l^{(t)}):=\frac{1}{S}\sum_{s=1}^S h_{\lambda^{(t)}}(\theta_s).\]
	  \item If $t\geq t_W$: compute the moving averaged lower bound
	  \[\overline {\text{LB}}_{t-t_W+1}=\frac{1}{t_W}\sum_{k=1}^{t_W} \wh{\text{LB}}(\l^{(t-k+1)}),\]
	  and if $\overline {\text{LB}}_{t-t_W+1}\geq\max(\overline\LB)$ patience = 0; else $\text{patience}:=\text{patience}+1$.
\item If $\text{patience}\geq P$, \texttt{stop=true}.
\item Set $t:=t+1$.
  \end{itemize}
\end{itemize}
\end{algorithm}

\begin{example}\label{exa:Example 2}
With the model and data in Example \ref{exa:Example 1}, let's derive a FFVB procedure for  
approximating the posterior $p(\mu,\s^2| y)\propto p(\mu)p(\s^2)p(y|\mu,\s^2)$ using Algorithm \ref{algorithm 2}. 
Suppose that the VB approximation is $q_\l(\mu,\sigma^2)=q(\mu)q(\sigma^2)$ with $q(\mu)=\N(\mu_\mu,\sigma_\mu^2)$ and $q(\sigma^2)=\text{Inverse-Gamma}(\a_{\sigma^2},\b_{\sigma^2})$.
This toy example is simply to demonstrate the use of Algorithm \ref{algorithm 2}, we do not
focus on the approximation accuracy here.

The model parameter is $\theta=(\mu,\sigma^2)^\top$ and the variational parameter $\lambda=(\mu_\mu,\sigma_\mu^2,\a_{\sigma^2},\b_{\sigma^2})^\top$.
In order to implement Algorithm \ref{algorithm 2}, we need $h_\lambda(\theta)=h(\theta)-\log q_\l(\t)$ with 
\bean
h(\theta)&=&\log \big(p(\mu)p(\sigma^2)p(y|\mu,\sigma^2)\big)\\
&=&-\frac{n+1}{2}\log(2\pi)-\frac12\log(\sigma_0^2)-\frac{(\mu-\mu_0)^2}{2\sigma_0^2}+\alpha_0\log(\beta_0)-\log\Gamma(\a_0)-(\frac{n}{2}+\a_0+1)\log(\s^2)\\
&&\phantom{cccc}-\frac{\b_0}{\s^2}-\frac{1}{2\s^2}\sum_{i=1}^n(y_i-\mu)^2,\\
\log q_\l(\t)&=&\a_{\s^2}\log\b_{\s^2}-\log\Gamma(\a_{\s^2})-(\a_{\s^2}+1)\log\s^2-\frac{\b_{\s^2}}{\s^2}-\frac12\log(2\pi)-\frac12\log(\s_{\mu}^2)-\frac{(\mu-\mu_\mu)^2}{2\s_\mu^2},
\eean
and
\[\nabla_\l\log q_\l(\t)=\Big(\frac{\mu-\mu_\mu}{\s_\mu^2},-\frac{1}{2\s_\mu^2}+\frac{(\mu-\mu_\mu)^2}{2\s_\mu^4},\log\b_{\s^2}-\frac{\Gamma'(\a_{\s^2})}{\Gamma(\a_{\s^2})}-\log\s^2,\frac{\a_{\s^2}}{\b_{\s^2}}-\frac{1}{\s^2} \Big)^\top.\]
We are now ready to implement Algorithm \ref{algorithm 2}.
Figure \ref{fig:FFVB example 2} plots the estimate of the posterior densities together with the lower bound.
The Variational Bayes estimates appear to be quite close to the Gibbs sampling
estimates in this example, with some small discrepancy between them. These estimates can be improved with more advanced
variants of FFVB presented later.

\begin{figure}[h]
\centering
\includegraphics[width=1\columnwidth]{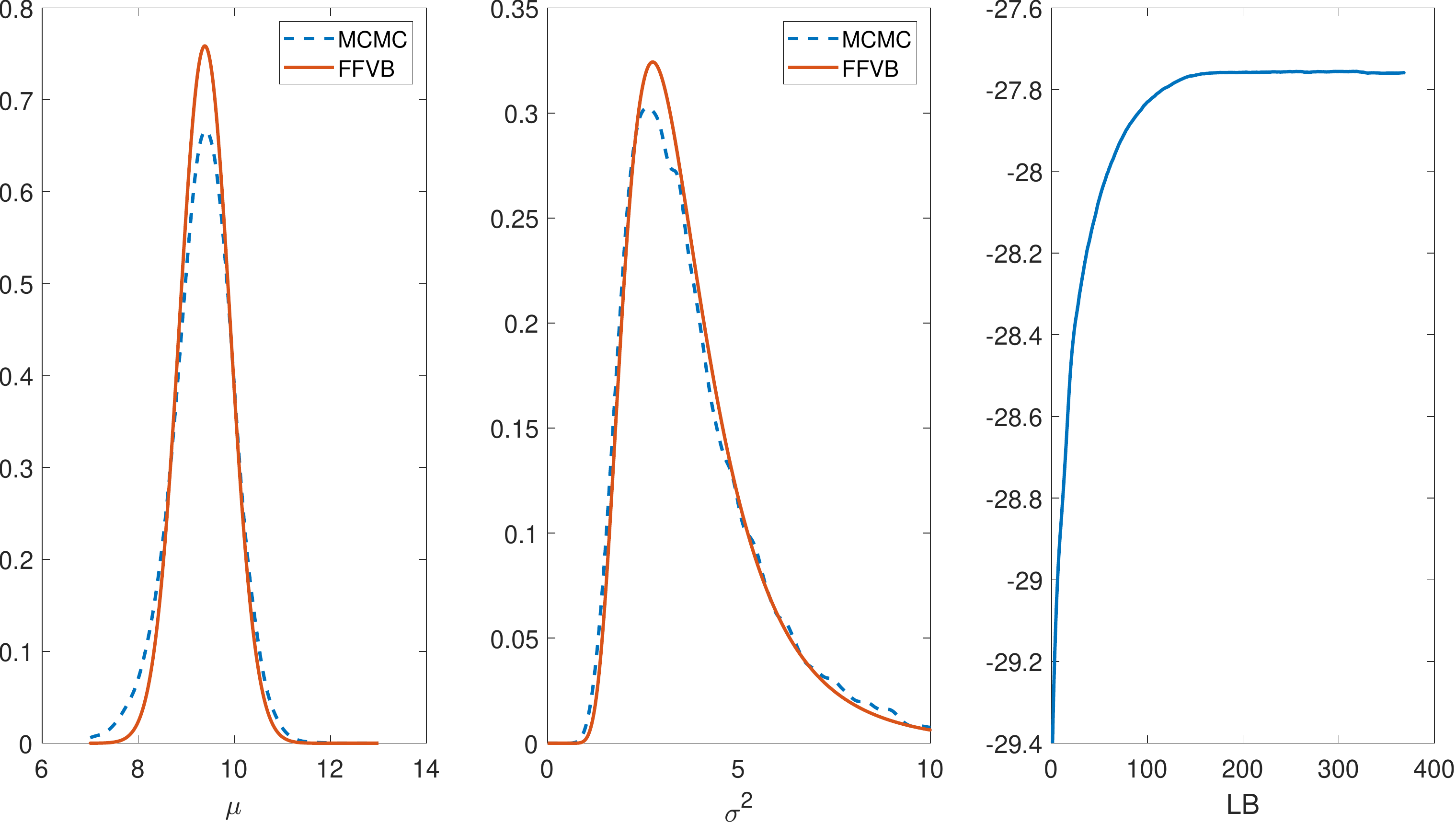}
\caption{Example \ref{exa:Example 2}: Posterior density for $\mu$ and $\sigma^2$ estimated by FFVB Algorithm \ref{algorithm 2} and Gibbs sampling. The last panel shows the smoothened lower bounds $\overline{\LB}_t$. The controlling parameters used are $S=2000$, $\beta_1=\beta_2=0.9$, $\eps_0=0.005$, $P=10$, $\tau=1000$ and $t_W=50$.} 
\label{fig:FFVB example 2}
\end{figure}
\end{example}

\begin{algorithm}[FFVB with control variates and natural gradient]\label{algorithm 3} 

{\bf Input}: Initial $\l^{(0)}$, momentum weight $\alpha_m$, fixed learning rate $\eps_0$, threshold $\tau$, rolling window size $t_W$ and maximum patience $P$. {\bf Model-specific requirement}: function $h(\theta):=\log\big(p(\theta)p(y|\theta)\big)$.
\begin{itemize}
  \item Initialization
  \begin{itemize}
	  \item Generate $\theta_s\sim q_{\lambda^{(0)}}(\theta)$, $s=1,...,S$.
	  \item Compute the unbiased estimate of the LB gradient
	  \[\wh{\nabla_\l\text{LB}}(\l^{(0)}):=\frac{1}{S}\sum_{s=1}^S\nabla_\lambda \log q_\lambda(\theta_s)\times h_\lambda(\theta_s)|_{\lambda=\lambda^{(0)}}\]
	  and the natural gradient 
	  \[\wh{\nabla_{\lambda}\LB} (\l^{(0)})^{\text{nat}} := I_F^{-1}(\l^{(0)})\wh{\nabla_\l\LB}(\l^{(0)}).\]
	  \item Set momentum gradient $\overline{{\nabla_\l{\LB}}}:=\wh{\nabla_{\lambda}\LB} (\l^{(0)})^{\text{nat}}$.
	  \item Estimate control variate vector $c$ as in \eqref{eq:optimal c_i} using the samples $\{\theta_s,s=1,...,S\}$.
	  \item Set $t=0$, $\text{patience} =0$ and \texttt{stop=false}.
  \end{itemize}
  \item While \texttt{stop=false}:
  \begin{itemize}
	  \item Generate $\theta_s\sim q_{\lambda^{(t)}}(\theta)$, $s=1,...,S$.
	  \item Compute the unbiased estimate of the LB gradient
	  \[\wh{\nabla_\l\text{LB}}(\l^{(t)})=\frac{1}{S}\sum_{s=1}^S\nabla_\lambda \log q_\lambda(\theta_s)\circ \big(h_\lambda(\theta_s)-c\big )|_{\lambda=\lambda^{(t)}}\]
 	  and the natural gradient 
	  \[\wh{\nabla_{\lambda}\LB} (\l^{(t)})^{\text{nat}} = I_F^{-1}(\l^{(t)})\wh{\nabla_\l\LB}(\l^{(t)}).\]
	  \item Estimate the new control variate vector $c$ as in \eqref{eq:optimal c_i} using the samples $\{\theta_s,s=1,...,S\}$.
	  \item Compute the momentum gradient
	  \[\overline{{\nabla_\l{\LB}}} = \alpha_\text{m} \overline{{\nabla_\l{\LB}}}+(1-\alpha_\text{m})\wh{\nabla_{\lambda}\LB}(\l^{(t)})^{\text{nat}}.\]
	  \item Compute $\alpha_t=\min(\epsilon_0,\epsilon_0\frac{\tau}{t})$ and update 
	  \[\l^{(t+1)}=\l^{(t)}+\a_t \overline{{\nabla_\l{\LB}}}.\]
	  \item Compute the lower bound estimate
	  \[\wh{\text{LB}}(\l^{(t)}):=\frac{1}{S}\sum_{s=1}^S h_{\lambda^{(t)}}(\theta_s).\]
	  \item If $t\geq t_W$: compute the moving average lower bound
	  \[\overline {\text{LB}}_{t-t_W+1}=\frac{1}{t_W}\sum_{k=1}^{t_W} \wh{\text{LB}}(\l^{(t-k+1)}),\]
	  and if $\overline {\text{LB}}_{t-t_W+1}\geq\max(\overline\LB)$ patience = 0; else $\text{patience}:=\text{patience}+1$.
\item If $\text{patience}\geq P$, \texttt{stop=true}.
\item Set $t:=t+1$.
  \end{itemize}
\end{itemize}
\end{algorithm}

\begin{example}\label{exa:Example 3}
With the model and data in Example \ref{exa:Example 1}, let's derive a FFVB procedure for  
approximating the posterior $p(\mu,\s^2| y)\propto p(\mu)p(\s^2)p( y|\mu,\s^2)$ using Algorithm \ref{algorithm 3}. 
In order to implement Algorithm \ref{algorithm 3}, apart from $h_\l(\t)$ and $\nabla_\l\log q_\l(\t)$ as in Example \ref{exa:Example 2}, we need the Fisher information matrix $I_F$.
It can be seen that this is a diagonal block matrix with two main blocks
\[
\begin{pmatrix}\frac{1}{\s_\mu^2}& 0\\
0&\frac{1}{2\s_\mu^4} \end{pmatrix},\;\;\;\text{ and }\;\;\;
\begin{pmatrix}
\frac{\partial^2\log\Gamma(\a_{\s^2})}{\partial \a_{\s^2}\partial \a_{\s^2}}&-\frac{1}{\b_{\s^2}}\\
-\frac{1}{\b_{\s^2}} & \frac{\a_{\s^2}}{\b_{\s^2}^2}
\end{pmatrix}.
\]

Figure \ref{fig:FFVB example 3} shows the estimated densities together with the lower bound estimates. 
In this example, Algorithm \ref{algorithm 3} appears to produce a very similar approximation as in Algorithm \ref{algorithm 2}.

\begin{figure}[h]
\centering
\includegraphics[width=1\columnwidth]{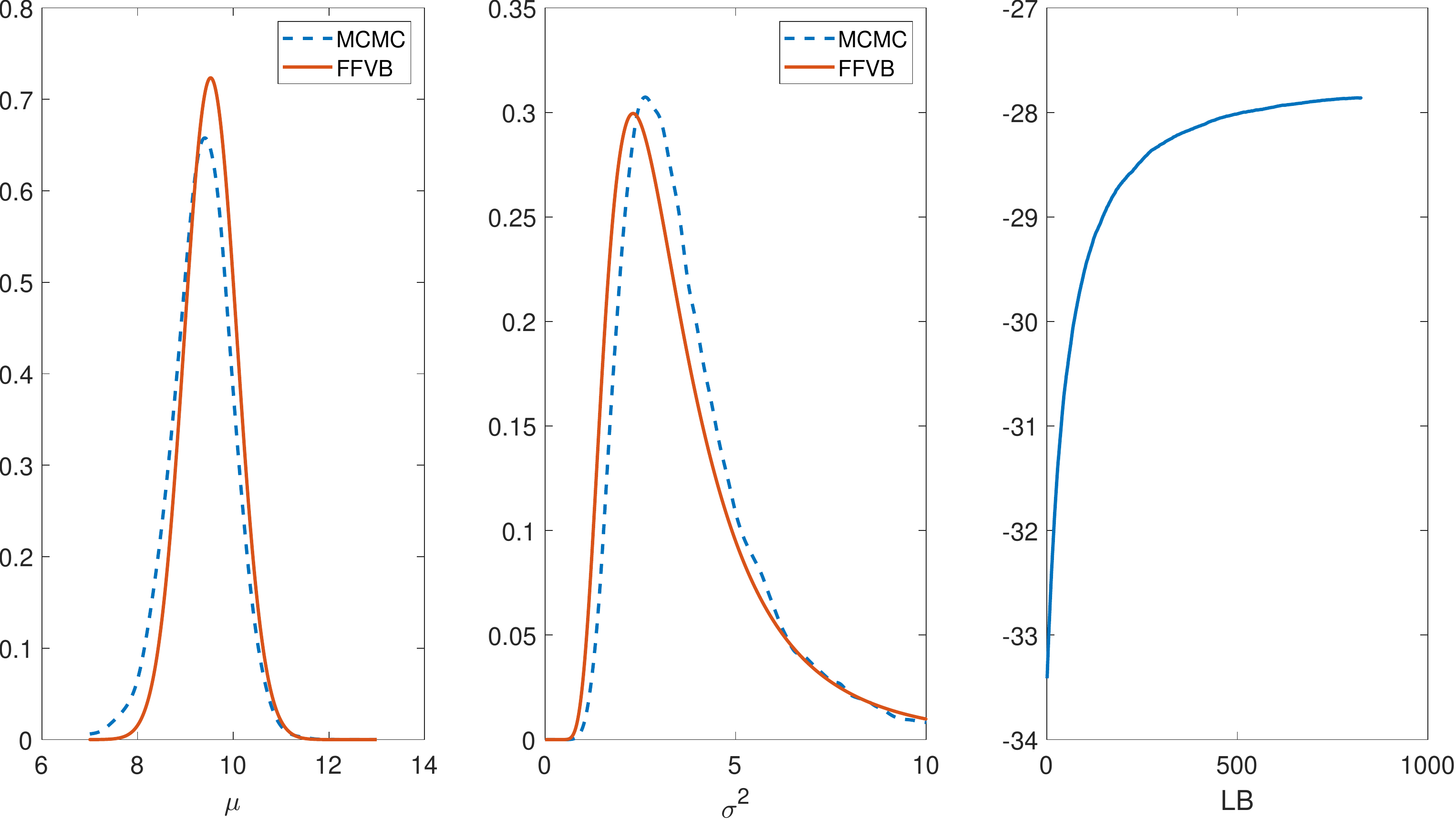}
\caption{Example \ref{exa:Example 3}: Posterior density for $\mu$ and $\sigma^2$ estimated by FFVB Algorithm \ref{algorithm 3} and Gibbs sampling. The last panel shows the averaged lower bounds $\overline{\LB}_t$.} 
\label{fig:FFVB example 3}
\end{figure}
\end{example}

The choice of the variational distribution $q_\l(\mu,\sigma^2)=q(\mu)q(\sigma^2)$ in Examples \ref{exa:Example 2} and \ref{exa:Example 3}
ignores the posterior dependence between $\mu$ and $\sigma^2$.
There are several alternatives that can improve this. 
One of these is to use Gaussian VB (see Section \ref{sec:GVB}) to approximate the posterior of the transformed parameter $\theta=\big(\mu,\log(\sigma^2)\big)$.
Another alternative is presented in Example \ref{exa:ChapterFFVB:example_hybrid} below.

\begin{example}\label{exa:ChapterFFVB:example_hybrid}
Consider again the model and data in Example \ref{exa:Example 1}.
It is possible to exploit the structure of this model to develop a better VB approximation.
Let us derive a FFVB procedure for approximating the posterior $p(\mu,\s^2|y)\propto p(\mu)p(\s^2)p(y|\mu,\s^2)$ using 
the variational distribution with density of the form
\beq\label{eq:ChapterFFVB:hybrid VB}
q_\lambda(\mu,\sigma^2)=\wt q_\lambda(\mu)p(\sigma^2|y,\mu),\;\;\;\wt q_\lambda(\mu)=\N(\mu_\mu,\sigma_\mu^2).
\eeq
This distribution, as the joint distribution of $\mu$ and $\sigma^2$, doesn't have a standard form, however, it is straightforward to sample from it.
This variational distribution exploits the standard form of the full conditional $p(\sigma^2|y,\mu)$, which is inverse-Gamma, and takes into account the posterior dependence between $\mu$ and $\sigma^2$.
 
The variational parameter $\lambda$ now only consists of $\mu_\mu$ and $\sigma_\mu^2$.
Using \eqref{eq:lb gradient}, the gradient of the lower bound is
\beqn
\nabla_\lambda\LB(\lambda)=\E_{q_\lambda(\mu,\sigma^2)}\Big(\nabla_\lambda\log \wt q_\lambda(\mu)\times h_\lambda(\theta)\Big)
\eeqn
with 
\beqn
h_\lambda(\theta)=\log p(\mu,\sigma^2)+\log p(y|\mu,\sigma^2)-\log \wt q_\lambda(\mu)-\log p(\sigma^2|y,\mu).
\eeqn
Algorithm \ref{algorithm 2} or Algorithm \ref{algorithm 3} now can be applied.

Figure \ref{fig:ChapterFFVB_FFVB:Example4_1_3} shows the estimated results.
As shown, this ``hybrid'' VB approximation is highly accurate in terms of both marginal density estimate and the joint density estimate.
%The hybrid VB method is discussed in detail in Section \ref{sec:ChapterFFVB:hybrid FFVB}.

\begin{figure}[h]
\centering
\includegraphics[width=1\textwidth,height=.4\textheight]{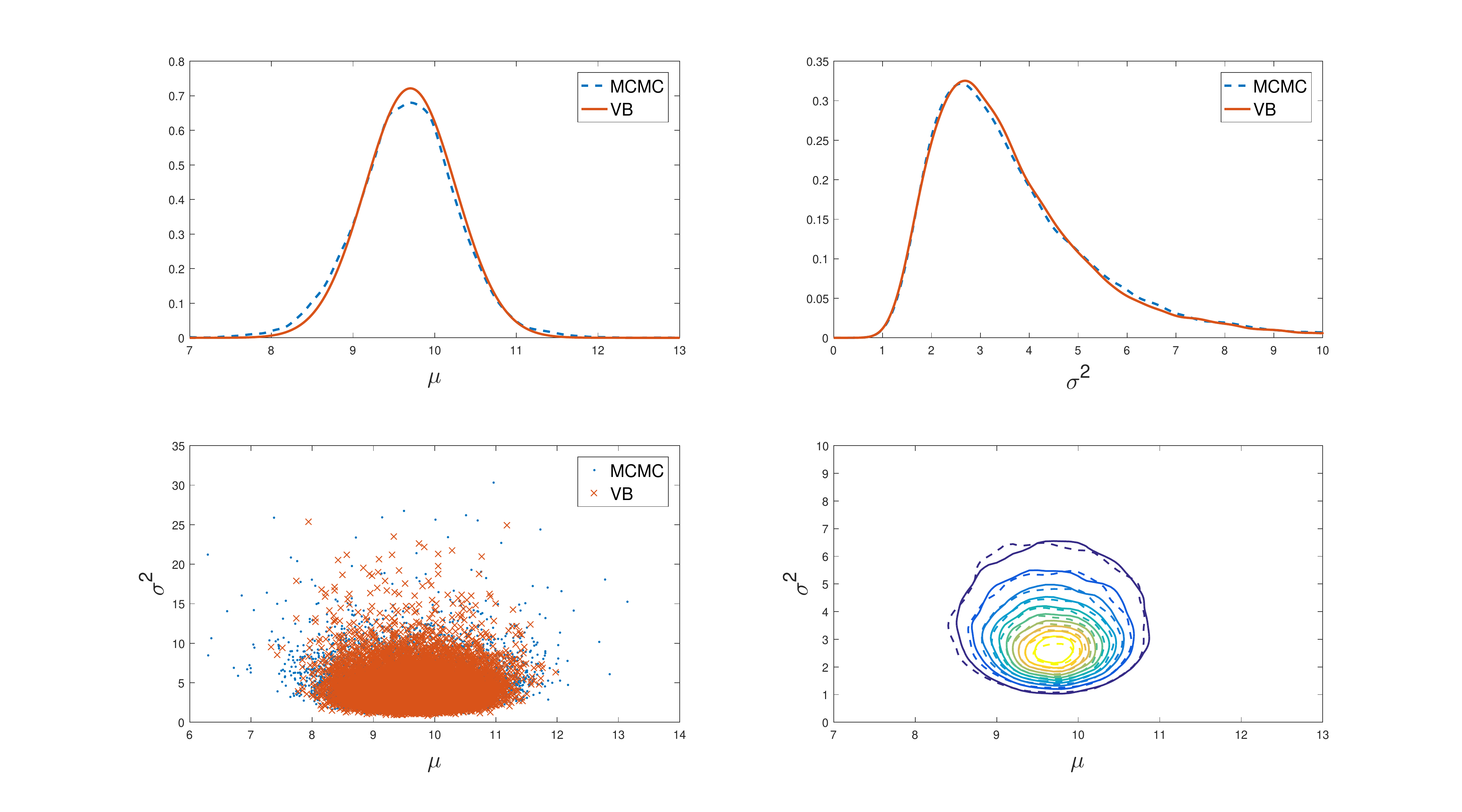}
\caption{Example \ref{exa:ChapterFFVB:example_hybrid}: First row: Posterior densities for $\mu$ and $\sigma^2$ estimated by Gibbs sampling and 
the hybrid VB method in \eqref{eq:ChapterFFVB:hybrid VB}. Second row: The joint samples and contour plot 
for the estimated joint posterior. In the bottom-right corner plot, the dashed lines are contours estimated based on the Gibbs samples,
and the solid lines estimated based on the samples generated from \eqref{eq:ChapterFFVB:hybrid VB}.}
\label{fig:ChapterFFVB_FFVB:Example4_1_3}
\end{figure}
\end{example}

%-------------------------------------------------------------%
\subsection{Reparameterization trick}\label{reparam trick}
%-------------------------------------------------------------%
The reparameterization trick is an attractive alternative to the control variate in Section \ref{sec:control variate}.
Suppose that for $\t\sim q_\l(\cdot)$, there exists a deterministic function $g(\l,\veps)$ such that $\t=g(\l,\veps)\sim q_\l(\cdot)$ where $\veps\sim p_\veps(\cdot)$.
We emphasize that $p_\veps(\cdot)$ must not depend on $\l$.
For example, if $q_\l(\t)=\N(\t;\mu,\sigma^2)$ then $\t=\mu+\sigma \veps$ with $\veps\sim\N(0,1)$. 
Writing $\LB(\l)$ as an expectation with respect to $p_\varepsilon(\cdot)$
\begin{align*}
  \LB(\l) & = \E_{\veps\sim p_\veps}\Big(h_\lambda(g(\veps,\l) )\Big),
\end{align*}
where $\E_{\veps\sim p_\veps}(\cdot)$ denotes expectation with respect to $p_\veps(\cdot)$, and differentiating under the integral sign gives
\begin{align}
  \nabla_\l\LB(\l) & = \E_{\veps\sim p_\veps}\Big(\nabla_\l g(\l,\veps)^\top\nabla_\theta h_\l (\theta)\Big) + \E_{\veps\sim p_\veps}\Big(\nabla_\lambda h_\l(\theta)\Big)\notag
\end{align}
where the $\theta$ within $h_\l(\theta)$ is understood as $\theta=g(\veps,\lambda)$ with $\lambda$ fixed. 
In particular, the gradient $\nabla_\lambda h_\l(\theta)$ is taken when $\theta$ is not considered as a function of $\lambda$.
Here, with some abuse of notation, $\nabla_\l g(\l,\veps)$ denotes the Jacobian matrix of size $d_\theta\times d_\lambda$ of the vector-valued function $\theta=g(\lambda,\veps)$. 
Note that
\begin{align*}
\E_{\veps\sim p_\veps}\Big(\nabla_\lambda h_\l(\theta)\Big)=\E_{\veps\sim q_\veps}\Big(\nabla_\lambda h_\l\big(\theta=g(\veps,\l)\big)\Big)&=-\E_{\veps\sim q_\veps}\Big(\nabla_\lambda \log q_\l\big(\theta=g(\veps,\l)\big)\Big)\\
&=-\E_{\theta\sim q_\l}\Big(\nabla_\lambda \log q_\l(\theta)\Big)=0,
\end{align*}
hence
\begin{equation}\label{eq:Reparameterization trick gradient}
  \nabla_\l\LB(\l) = \E_{\veps\sim q_\veps}\Big(\nabla_\l g(\l,\veps)^\top\nabla_\theta h_\l (\theta)\Big).
\end{equation}
The gradient \eqref{eq:Reparameterization trick gradient} can be estimated unbiasedly using  i.i.d samples $\veps_s\sim p_\veps(\cdot)$, $s=1,...,S$, as
\begin{align}
\wh{\nabla_\l{\LB}}(\l) & =\frac1S\sum_{s=1}^S\nabla_\l g(\l,\veps_s)^\top \nabla_\t \big\{h_\lambda(g(\l,\veps_s))\big\}. \label{roedergradest}
\end{align}
The {\it reparametrization gradient} estimator \eqref{roedergradest} is often more efficient than alternative approaches to estimating
the lower bound gradient, partly because it takes into account the information from the gradient $\nabla_\theta h_\l(\theta)$. 
In typical VB applications, the number of Monte Carlo samples $S$ used in estimating the lower bound gradient can be as small as 5 if 
the reparameterization trick is used, while the control variates method requires an $S$ of about hundreds or more.
However, there is a dilemma about choosing $S$ that we must be careful of.
With the reparameterization trick, a small $S$ might be enough for estimating the lower bound gradient,
however, we still need a moderate $S$ in order to obtain a good estimate of the lower bound if lower bound is used in the stopping criterion.
Also, compared to score-function gradient, FFVB approaches that use reparameterization gradient require not only the model-specific function $h(\theta)$ but also its gradient $\nabla_\theta h(\theta)$.

Algorithm \ref{algorithm 4} provides a detailed implementation of the FFVB approach that uses the reparameterization trick and adaptive learning.
A small modification of Algorithm \ref{algorithm 4} (not presented) gives the implementation of the FFVB approach that uses the reparameterization trick and natural gradient.

\begin{algorithm}[FFVB with reparameterization trick and adaptive learning]\label{algorithm 4} 
{\bf Input}: Initial $\l^{(0)}$, adaptive learning weights $\beta_1,\beta_2\in(0,1)$, fixed learning rate $\eps_0$, threshold $\tau$, rolling window size $t_W$ and maximum patience $P$. {\bf Model-specific requirement}: function $h(\theta):=\log\big(p(\theta)p(y|\theta)\big)$ and its gradient $\nabla_\theta h(\theta)$.
\begin{itemize}
  \item Initialization
  \begin{itemize}
	  \item Generate $\varepsilon_s\sim p_{\veps}(\cdot)$, $s=1,...,S$.
	  \item Compute the unbiased estimate of the LB gradient
	  \[\wh{\nabla_\l\text{LB}}(\l^{(0)}):=\frac1S\sum_{s=1}^S\nabla_\l g(\l,\veps_s)^\top \nabla_\t \big\{h_\lambda(g(\l,\veps_s))\big\}\big|_{\lambda=\lambda^{(0)}}.\]
	  \item Set $g_0:=\wh{\nabla_\l\text{LB}}(\l^{(0)})$, $v_0:=(g_0)^2$, $\bar g:=g_0$, $\bar v:=v_0$. 
	  \item Set $t=0$, $\text{patience}=0$ and \texttt{stop=false}.
  \end{itemize}
  \item While \texttt{stop=false}:
  \begin{itemize}
	  \item Generate $\veps_s\sim p_{\veps}(\cdot)$, $s=1,...,S$
	  \item Compute the unbiased estimate of the LB gradient
\[g_t:=\wh{\nabla_\l\text{LB}}(\l^{(t)})=\frac1S\sum_{s=1}^S\nabla_\l g(\l,\veps_s)^\top \nabla_\t \big\{h_\lambda(g(\l,\veps_s))\big\}\big|_{\lambda=\lambda^{(t)}}.\]
	  \item Compute $v_t=(g_t)^2$ and 
	  \[\bar g =\beta_1 \bar g+(1-\beta_1)g_t,\;\;\bar v =\beta_2 \bar v+(1-\beta_2)v_t.\]
	  \item Compute $\alpha_t=\min(\varepsilon_0,\varepsilon_0\frac{\tau}{t})$ and update
	  \[\l^{(t+1)}=\l^{(t)}+\a_t \bar g/\sqrt{\bar v}\]
	  \item Compute the lower bound estimate
	  \[\wh{\text{LB}}(\l^{(t)}):=\frac{1}{S}\sum_{s=1}^S h_{\lambda^{(t)}}(\theta_s),\;\;\;\theta_s=g(\lambda^{(t)},\veps_s).\]
	  \item If $t\geq t_W$: compute the moving average lower bound
	  \[\overline {\text{LB}}_{t-t_W+1}=\frac{1}{t_W}\sum_{k=1}^{t_W} \wh{\text{LB}}(\l^{(t-k+1)}),\]
	  and if $\overline {\text{LB}}_{t-t_W+1}\geq\max(\overline\LB)$ patience = 0; else $\text{patience}:=\text{patience}+1$.
\item If $\text{patience}\geq P$, \texttt{stop=true}.
\item Set $t:=t+1$.
  \end{itemize}
\end{itemize}
\end{algorithm}

\subsection{Gaussian Variational Bayes}\label{sec:GVB}
The most popular VB approaches are probably Gaussian VB where the approximation $q_\l(\theta)$ is a Gaussian distribution with mean $\mu$ and covariance matrix $\Sigma$. This section presents several variants of this GVB approach.

\subsubsection{GVB with Cholesky decomposed covariance}\label{sec: GVB Cholesky}
This GVB method uses the Cholesky decomposition for the covariance matrix $\Sigma$, $\Sigma=LL^\top$ with $L$ a lower triangular matrix\footnote{For the Cholesky decomposition of $\Sigma$ to be unique, one needs the constraint that the diagonal entries of $L$ to be strictly positive. For simplicity, however, we do not impose this constraint here.}. 
We will use the reparameterization trick for variance reduction.
A sample $\theta\sim q_\lambda(\theta)$ can be written as $\theta=g(\lambda,\veps)=\mu+L\veps$ with $\veps\sim \N_d(0,I_d)$, and $d$ the dimension of $\theta$. The variational parameter vector $\lambda$ includes $\mu$ and the non-zero elements of $L$. 
As Jacobian matrix $\nabla_\mu g(\lambda,\veps)=I$, the identity matrix, from \eqref{eq:Reparameterization trick gradient},
the gradient of the lower bound w.r.t. $\mu$ is
\beqn
\nabla_\mu\LB(\l)=\E_\veps\big[\nabla_\theta h_\lambda(\theta)\big],\;\;\;\text{with}\;\;\;\theta=\mu+L\veps.
\eeqn
To compute the gradient w.r.t. $L$, we first need some notations.
For a $d\times d$ matrix $A$, denote by $\vec(A)$
the $d^2$-vector obtained by stacking the columns of $A$ from left to right one underneath the other,
by $\vech(A)$ the $\frac12d(d+1)$-vector obtained by stacking the columns of the lower triangular part of $A$,
and by $A\otimes B$ the Kronecker product of matrices $A$ and $B$.
For any matrices $A$, $B$ and $X$ of suitable sizes, we shall use the fact that $\text{vec}(AXB)=(B^\top\otimes A)\text{vec}(X)$.
Then, $L\veps=\text{vec}(I_dL\veps)=(\veps^\top\otimes I_d)\text{vec}(L)$ and hence $\nabla_{\text{vec}(L)} g(\lambda,\veps)=\veps^\top\otimes I_d$.
From \eqref{eq:Reparameterization trick gradient},
\bean
\nabla_{\text{vec}(L)}\LB(\lambda)&=&\E_{\veps}\Big[\nabla_{\text{vec}(L)}g(\lambda,\veps)^\top\nabla_\theta h_\lambda(\theta)\Big]\\
&=&\E_{\veps}\Big[(\veps\otimes I_d) \nabla_{\theta} h_\lambda(\theta)\Big]\\
&=&\E_{\veps}\Big[\text{vec}\big(\nabla_{\theta} h_\lambda(\theta)\veps^\top\big)\Big],\;\;\;\text{with}\;\;\;\theta=\mu+L\veps.
\eean
This implies that
\beq
\nabla_{\vech(L)}\LB(\l)=\E_{\veps}\big[\vech\big(\nabla_{\theta} h_\lambda(\theta)\veps^\top\big)\big].
\eeq
From Algorithm \ref{algorithm 4}, we arrive at the following GVB algorithm, referred to below as Cholesky GVB.

\begin{algorithm}[Cholesky GVB]\label{alg:GVB-Cholesky} 
{\bf Input}: Initial $\mu^{(0)}$, $L^{(0)}$ and $\l^{(0)}:=({\mu^{(0)}}^\top,\vech(L^{(0)})^\top)^\top$, number of samples $S$, adaptive learning weights $\beta_1,\beta_2\in(0,1)$, fixed learning rate $\eps_0$, threshold $\tau$, rolling window size $t_W$ and maximum patience $P$. {\bf Model-specific requirement}: function $h(\theta)$ and $\nabla_\theta h(\theta)$.
\begin{itemize}
  \item Initialization
  \begin{itemize}
	  \item Generate $\varepsilon_s\sim N_d(0,I)$, $s=1,...,S$.
	  \item Compute the estimate of the lower bound gradient\\
	  $\wh{\nabla}_\l\LB(\l^{(0)})=(\wh{\nabla}_\mu\LB(\l^{(0)})^\top,\wh{\nabla}_{\vech(L)}\LB(\l^{(0)})^\top)^\top$ where
	  \bean
	  \wh{\nabla}_\mu\LB(\l^{(0)})&:=&\frac1S\sum_{s=1}^S\nabla_\theta h_\lambda(\theta_s),\\
	  \wh{\nabla}_{\vech(L)}\LB(\l^{(0)})&:=&\frac1S\sum_{s=1}^S\vech\big(\nabla_\theta h_\lambda(\theta_s)\veps_s^\top\big),
	  \eean
	  with $\theta_s=\mu^{(0)}+L^{(0)}\veps_s$.
	  \item Set $g_0:=\wh{\nabla}_\l\mathcal{L}(\l^{(0)})$, $v_0:=(g_0)^2$, $\bar g:=g_0$, $\bar v:=v_0$. 
	  \item Set $t=0$, $\text{patience}=0$ and \texttt{stop=false}.
  \end{itemize}
  \item While \texttt{stop=false}:
  \begin{itemize}
	  \item Generate $\veps_s\sim p_{\veps}(\cdot)$, $s=1,...,S$. Recalculate $\mu^{(t)}$ and $L^{(t)}$ from $\lambda^{(t)}$.
  	  \item Compute the estimate of the lower bound gradient\\ $g_t:=\wh{\nabla}_\l\LB(\l^{(t)})=(\wh{\nabla}_\mu\LB(\l^{(t)})^\top,\wh{\nabla}_{\vech(L)}\LB(\l^{(t)})^\top)^\top$ where
	  \bean
	  \wh{\nabla}_\mu\LB(\l^{(t)})&:=&\frac1S\sum_{s=1}^S\nabla_\theta h_\lambda(\theta_s),\\
	  \wh{\nabla}_{\vech(L)}\LB(\l^{(t)})&:=&\frac1S\sum_{s=1}^S\vech\big(\nabla_\theta h_\lambda(\theta_s)\veps_s^\top\big),
	  \eean
	  with $\theta_s=\mu^{(t)}+L^{(t)}\veps_s$.
	  \item Compute $v_t=(g_t)^2$ and 
	  \[\bar g =\beta_1 \bar g+(1-\beta_1)g_t,\;\;\bar v =\beta_2 \bar v+(1-\beta_2)v_t.\]
	  \item Compute $\alpha_t=\min(\varepsilon_0,\varepsilon_0\frac{\tau}{t})$ and update
	  \[\l^{(t+1)}=\l^{(t)}+\a_t \bar g/\sqrt{\bar v}\]
	  \item Compute the lower bound estimate
	  \[\wh{\mathcal{L}}(\l^{(t)}):=\frac{1}{S}\sum_{s=1}^S h_\lambda(\theta_s).\]
	  \item If $t\geq t_W$: compute the moving averaged lower bound
	  \[\overline {\mathcal{L}}_{t-t_W+1}=\frac{1}{t_W}\sum_{k=1}^{t_W} \wh{\mathcal{L}}(\l^{(t-k+1)}),\]
	  and if $\overline {\mathcal{L}}_{t-t_W+1}\geq\max(\overline\LB)$ patience = 0; else $\text{patience}:=\text{patience}+1$.
\item If $\text{patience}\geq P$, \texttt{stop=true}.
\item Set $t:=t+1$.
  \end{itemize}
\end{itemize}
\end{algorithm}

\begin{example}[Bayesian logistic regression]\label{exa:Example 4}
Consider a Bayesian logistic regression problem with design matrix $X=[x_1,...,x_n]^\top$ and vector of binary responses $y$.
The log-likelihood is
\[\log p(y|X,\theta)=y^\top X\theta-\sum_{i=1}^n\log\big(1+\exp(x_i^\top\theta)\big)\]
with $\theta$ the vector of $d$ coefficients. Suppose that a normal prior $\N(0,\sigma_0^2I)$ is used for $\theta$.
To implement the Cholesky GVB method, all we need is the function 
\beq
h(\theta)=\log p(\theta)+\log p(y|X,\theta)=-\frac{d}{2}\log(2\pi)-\frac{d}{2}\log(\sigma_0^2)-\frac{\theta^\top\theta}{2\sigma_0^2}+y^\top X\theta-\sum_{i=1}^n\log\big(1+\exp(x_i^\top\theta)\big),
\label{eqn:h_function_ex4}
\eeq
and its gradient
\beq
\nabla_\theta h(\theta)=-\frac{1}{\sigma_0^2}\theta+X^\top\big(y-\pi(\theta)\big)
\label{eqn:grad_h_function_ex4_1}
\eeq
with
\beq
\pi(\theta)=\Big(\frac{1}{1+\exp(-x_1^\top\theta)},\cdots,\frac{1}{1+\exp(-x_n^\top\theta)}\Big)^\top.
\label{eqn:grad_h_function_ex4_2}
\eeq
The Labour Force Participation dataset contains information of 753 women with one binary variable indicating whether or not they are currently in the labour force
together with seven covariates such as number of children under 6 years old, age, education level, etc.
Figure \ref{fig:GVB_Cholesky} plots the VB approximation for each coefficient $\theta_i$ together with the lower bound estimates over the iterations.

\begin{figure}[h]
\centering
\includegraphics[width=1\columnwidth]{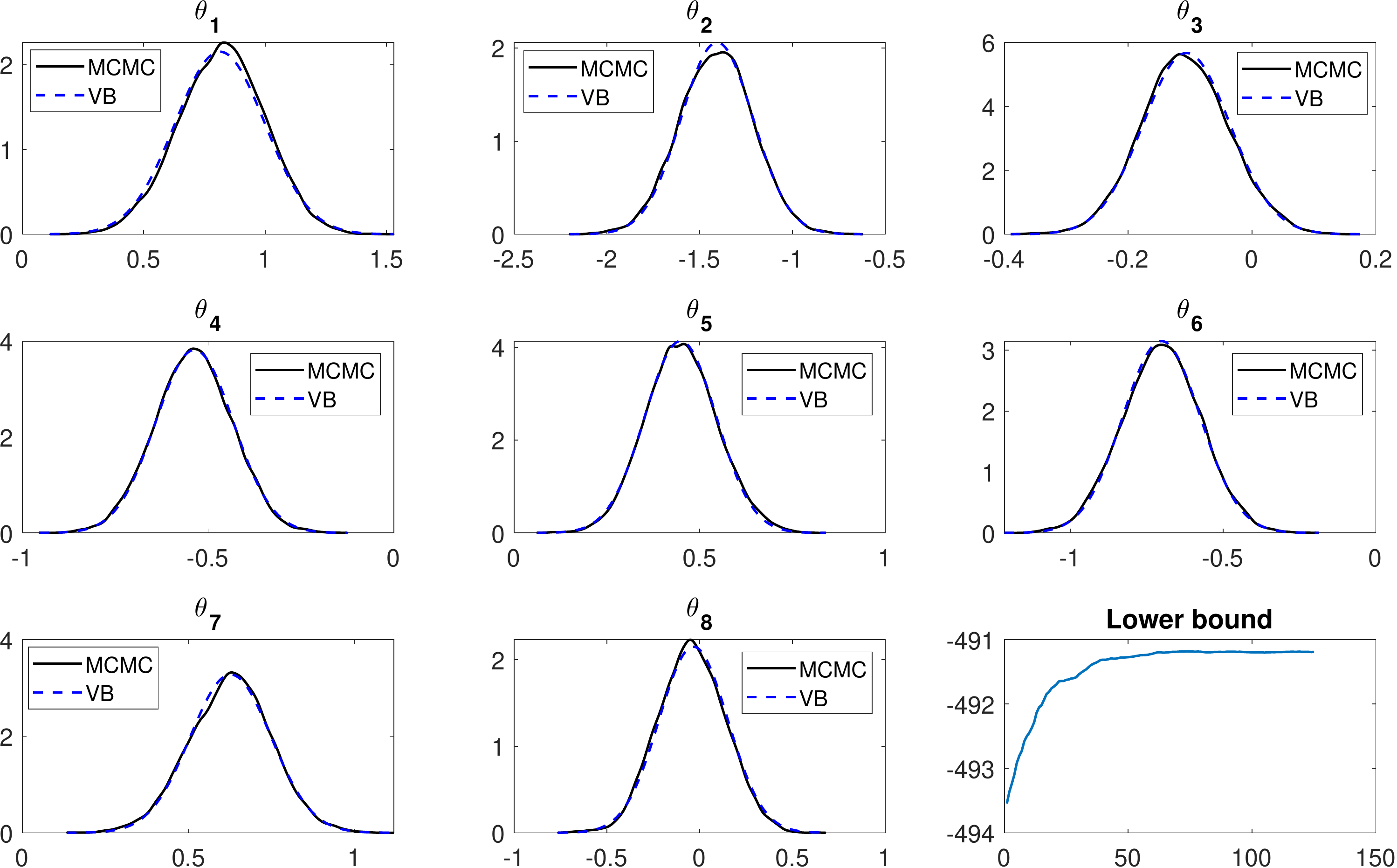}
\caption{Cholesky GVB for approximating the posterior in logistic regression. The CPU time was roughly 3 seconds.
The controlling parameters used are $S=50$, $\beta_1=\beta_2=0.9$, $\eps_0=0.002$, $P=20$, $\tau=500$ and $t_W=50$.
}
\label{fig:GVB_Cholesky}
\end{figure}
\end{example}

\subsubsection{GVB with factor decomposed covariance}\label{sec: GVB factor}
An alternative to the Cholesky decomposition is the factor decomposition
\[\Sigma=BB^\top+C^2,\]
where $B$ is the factor loading matrix of size $d\times f$ with $f\ll d$ the number of factors and $C$ is a diagonal matrix, $C=\diag(c_1,...,c_d)$.
GVB with this factor covariance structure is useful in high-dimensional settings where $d$ is large,
as the number of variational parameters reduces from $d+d*(d+1)/2$ in the case of full Gaussian to $(f+2)d$ in the case of factor decomposition.
This VB method is first developed in \cite{Ong:2018} who term the method Variational Approximation with Factor Covariance (VAFC)
and use Algorithm \ref{algorithm 4} for training, as computing the natural gradient in this case is difficult.
This section describes the case with one factor, $f=1$, which achieves a great computational speed-up for approximate Bayesian inference in big models such as deep neural networks where $d$ can be very large.
Also, with $f=1$, \cite{Tran:2020} show that it is possible to calculate the natural gradient efficiently  
and term their method NAtural gradient Gaussian Variational Approximation with factor Covariance (NAGVAC).

With $f=1$, we rewrite the factor decomposition as 
\[\Sigma=bb^\top+C^2,\quad\;\;C=\diag(c),\]  
where $b=(b_1,...,b_d)^\top$ and $c=(c_1,...,c_d)^\top$ are vectors. The variational parameter vector is $\lambda=(\mu^\top,b^\top,c^\top)^\top$.
Using the reparameterization trick, $\theta\sim\N(\mu,\Sigma)$ can be written as
\beqn
\theta = g(\lambda,\veps)=\mu+\veps_1b+c\circ\veps_2
\eeqn
where $\veps=(\veps_1,\veps_2^\top)^\top\sim\N_{d+1}(0,I)$, and $c\circ\veps_2$ denotes the component-wise product of vectors $c$ and $\veps_2$.
Note that
\[\nabla_\mu g(\lambda,\veps)=I_d,\;\;\;\;\nabla_b g(\lambda,\veps)=\veps_1I_d,\;\;\;\;\nabla_c g(\lambda,\veps) =\diag(\veps_2),\]
hence the reparameterization gradient is
\beq\label{eq: GVB factor grad est}
\nabla_\l\LB(\l)=\E_{q_\veps}\begin{pmatrix}\nabla_\theta h_\lambda(\mu+\veps_1b+c\circ\veps_2)\\
\veps_1 \nabla_\theta h_\lambda(\mu+\veps_1b+c\circ\veps_2)\\
\veps_2 \circ \nabla_\theta h_\lambda(\mu+\veps_1b+c\circ\veps_2)
\end{pmatrix}.
\eeq  
The gradient of function $h_\lambda(\theta)$ is
\[\nabla_\theta h_\lambda(\theta)=\nabla_\theta h(\theta)-\nabla_\theta\log q_\lambda(\theta)=\nabla_\theta\log\big(p(\theta)p(y|\theta)\big)-\nabla_\theta\log q_\lambda(\theta),\]
where the first term is model-specific and the second term is $\nabla_\theta\log q_\lambda(\theta)=-\Sigma^{-1}(\theta-\mu)$.
To avoid computing directly the inverse $\Sigma^{-1}$ and the matrix-vector multiplication, noting that $\Sigma^{-1}=C^{-2}-\frac{1}{1+b^\top C^{-2}b} C^{-2}bb^\top  C^{-2}$, we have
\beqn
\nabla_\theta\log q_\lambda(\theta)=-(\theta-\mu)\circ c^{-2}+\frac{(b\circ c^{-2})^\top(\theta-\mu)}{1+(b\circ c^{-1})^\top (b\circ c^{-1})}(b\circ c^{-2}),
\eeqn
with $c^{-1}:=(1/c_1,...,1/c_d)^\top$ and $c^{-2}:=(1/c_1^2,...,1/c_d^2)^\top$.
To compute lower bound estimates, we need
\beqn
\log q_\lambda(\theta) = -\frac{d}{2}\log(2\pi)-\frac12\log|\Sigma|-\frac12(\theta-\mu)^\top\Sigma^{-1}(\theta-\mu).
\eeqn
As $\Sigma=C\big((C^{-1}b)(C^{-1}b)^\top+I\big)C$,
\beqn
|\Sigma| = |C|^2\big(1+(C^{-1}b)^\top(C^{-1}b)\big)=\big(\prod_{i=1}^d{c_i^2}\big)\big(1+\sum_{i=1}^d\frac{b_i^2}{c_i^2}\big).
\eeqn
Hence, a computationally efficient version of $\log q_\lambda(\theta)$ is
\bean
\log q_\lambda(\theta) &=& -\frac{d}{2}\log(2\pi)-\frac12\sum_{i=1}^d\log c_i^2-\frac12\log\big(1+\sum_{i=1}^d\frac{b_i^2}{c_i^2}\big)\\
&&\phantom{ccc}-\frac12(\theta-\mu)^\top\big((\theta-\mu)\circ c^{-2}\big)+\frac{\big((b\circ c^{-2})^\top(\theta-\mu)\big)^2}{2\big(1+(b\circ c^{-1})^\top(b\circ c^{-1})\big)}.
\eean
Finally, it can be shown that the natural gradient in \eqref{eq:natural gradient} can be approximately computed in closed form 
as in the following algorithm (see \citet{Tran:2020}), whose computational complexity is $O(d)$.
\begin{algorithm}[Computing the natural gradient]\label{alg:GVB natural gradient}
Input: Vector $b$, $c$ and ordinary gradient of the lower bound $g = (g_1^\top,g_2^\top,g_3^\top)^\top$ with $g_1$ the vector formed by the first $d$ elements of $g$,
$g_2$ formed by the next $d$ elements, and $g_3$ the last $d$ elements. 
Output: The natural gradient $g^{\text{nat}}=I_F^{-1}g$.
\begin{itemize}
\item Compute the vectors $v_1=c^2-2b^2\circ c^{-4}$, $v_2=b^2\circ c^{-3}$, and the scalars $\kappa_1=\sum_{i=1}^db_i^2/c_i^2$, $\kappa_2=\frac{1}{2}(1+\sum_{i=1}^d v_{2i}^2/v_{1i})^{-1}$.
\item Compute 
\[g^{\text{nat}}=\begin{pmatrix}
(g_1^\top b)b+c^2\circ g_1\\
\frac{1+\kappa_1}{2\kappa_1}\Big((g_2^\top b)b+c^2\circ g_2\Big)\\
\frac12v_1^{-1}\circ g_3+\kappa_2 \big[(v_1^{-1}\circ v_2)^\top g_3\big](v_1^{-1}\circ v_2)
\end{pmatrix}.\]
\end{itemize}
\end{algorithm}
We now describe the NAGVAC algorithm that can be used as a fast VB method for approximate Bayesian inference 
in high-dimensional applications such as Bayesian deep neural networks.
In such applications, instead of using the lower bounds for stopping rule,
one often uses a loss function evaluated on a validation dataset for stopping.
Then, the updating is stopped if the loss function is not decreased after $P$ iterations.

\begin{algorithm}[NAGVAC]\label{alg:ChapterFFVB:GVB-factor} 

{\bf Input}: Initial $\lambda^{(0)}:=(\mu^{(0)},b^{(0)},c^{(0)})$, number of samples $S$, momentum weight $\a_m$, fixed learning rate $\eps_0$, threshold $\tau$, rolling window size $t_W$ and maximum patience $P$. {\bf Model-specific requirement}: function $h(\theta)$ and $\nabla_\theta h(\theta)$.
\begin{itemize}
  \item Initialization
  \begin{itemize}
	  \item Generate $\varepsilon_{1,s}\sim \N(0,1)$ and $\varepsilon_{2,s}\sim \N_d(0,I_d)$, $s=1,...,S$.
	  \item Compute the lower bound gradient estimate $\wh{\nabla}_\l\LB(\l^{(0)})$ as in \eqref{eq: GVB factor grad est}, and then compute the natural gradient $\wh{\nabla_{\lambda}\LB} (\l^{(0)})^{\text{nat}}$ using Algorithm \ref{alg:GVB natural gradient}.
	  \item Set momentum gradient $\overline{{\nabla_\l{\LB}}}:=\wh{\nabla_{\lambda}\LB} (\l^{(0)})^{\text{nat}}$.
	  \item Set $t=0$, $\text{patience}=0$ and \texttt{stop=false}.
  \end{itemize}
  \item While \texttt{stop=false}:
  \begin{itemize}
	  \item Generate $\varepsilon_{1,s}\sim \N(0,1)$ and $\varepsilon_{2,s}\sim \N_d(0,I_d)$, $s=1,...,S$.	  
	  \item Compute the lower bound gradient estimate $\wh{\nabla}_\l\LB(\l^{(t)})$ as in \eqref{eq: GVB factor grad est}, and then compute the natural gradient $\wh{\nabla_{\lambda}\LB} (\l^{(t)})^{\text{nat}}$ using Algorithm \ref{alg:GVB natural gradient}.
	  \item Compute the momentum gradient
	  \[\overline{{\nabla_\l{\LB}}} = \alpha_\text{m} \overline{{\nabla_\l{\LB}}}+(1-\alpha_\text{m})\wh{\nabla_{\lambda}\LB}(\l^{(t)})^{\text{nat}}.\]
	  \item Compute $\alpha_t=\min(\epsilon_0,\epsilon_0\frac{\tau}{t})$ and update 
	  \beqn
	  \l^{(t+1)}=\l^{(t)}+\a_t \overline{{\nabla_\l{\LB}}}.
	  \eeqn
	  \item Compute the validation loss $\text{Loss}(\lambda^{(t)})$. If $\text{Loss}(\lambda^{(t)})\leq \min\{\text{Loss}(\lambda^{(1)}),...,\text{Loss}(\lambda^{(t-1)})\}$ patience = 0; else $\text{patience}:=\text{patience}+1$.
%	  \item Compute the lower bound estimate
%	  \[\wh{\mathcal{L}}(\l^{(t)}):=\frac{1}{S}\sum_{s=1}^S h_{\lambda^{(t)}}(\theta_s).\]
%	  \item If $t\geq t_W$: compute the moving averaged lower bound
%	  \[\overline {\mathcal{L}}_{t-t_W+1}=\frac{1}{t_W}\sum_{k=1}^{t_W} \wh{\mathcal{L}}(\l^{(t-k+1)}),\]
%	  and if $\overline {\mathcal{L}}_{t-t_W+1}\geq\max(\overline\LB)$ patience = 0; else $\text{patience}:=\text{patience}+1$.
\item If $\text{patience}\geq P$, \texttt{stop=true}.
\item Set $t:=t+1$.
  \end{itemize}
\end{itemize}
\end{algorithm}

\begin{example}[Bayesian deep neural net]\label{exa:Bayesian deep neural net}
This example brieftly presents an application of the NAGVAC method for
fitting a Bayesian deep neural network (BNN). See \cite{Tran:2020} for a detailed description of this example.
Bayesian deep neural network models are an example of big models where the size $d$ of unknown parameters can be in thousands or millions.
We consider the census dataset extracted from the U.S. Census Bureau database and available on the UCI
Machine Learning Repository \url{https://archive.ics.uci.edu/ml/index.php}.
The prediction task is to determine whether a person's income is over \$50K per year,
based on 14 attributes including age, workclass, race, etc, of which many are categorical variables.
After using dummy variables to represent the categorical variables, there are 103 input variables.
The training dataset has 24,129 observations and the validation set has 6032 observations.
As is typical in Deep Learning applications, here we use the minus log-likelihood computed on the validation set
as the loss function to judge when to stop the VB training algorithm.
The structure of the neural net is $[104, 100, 100]$: input layer with 104 variables including the intercept, and two hidden layers each with 100 units.
The size of parameters $\theta$ is 20,500.
Algorithm \ref{alg:ChapterFFVB:GVB-factor} for training this deep learning model stopped after 2812 iterations.
Figure \ref{fig:ChapterFFVB_NAGVAG} plots the validation loss over the iterations.
For a detailed discussion on the prediction accuracy of this BNN compared to the Bayesian logistic model, see \cite{Tran:2020}.
  
\begin{figure}[h]
\centering
\includegraphics[width=1\columnwidth]{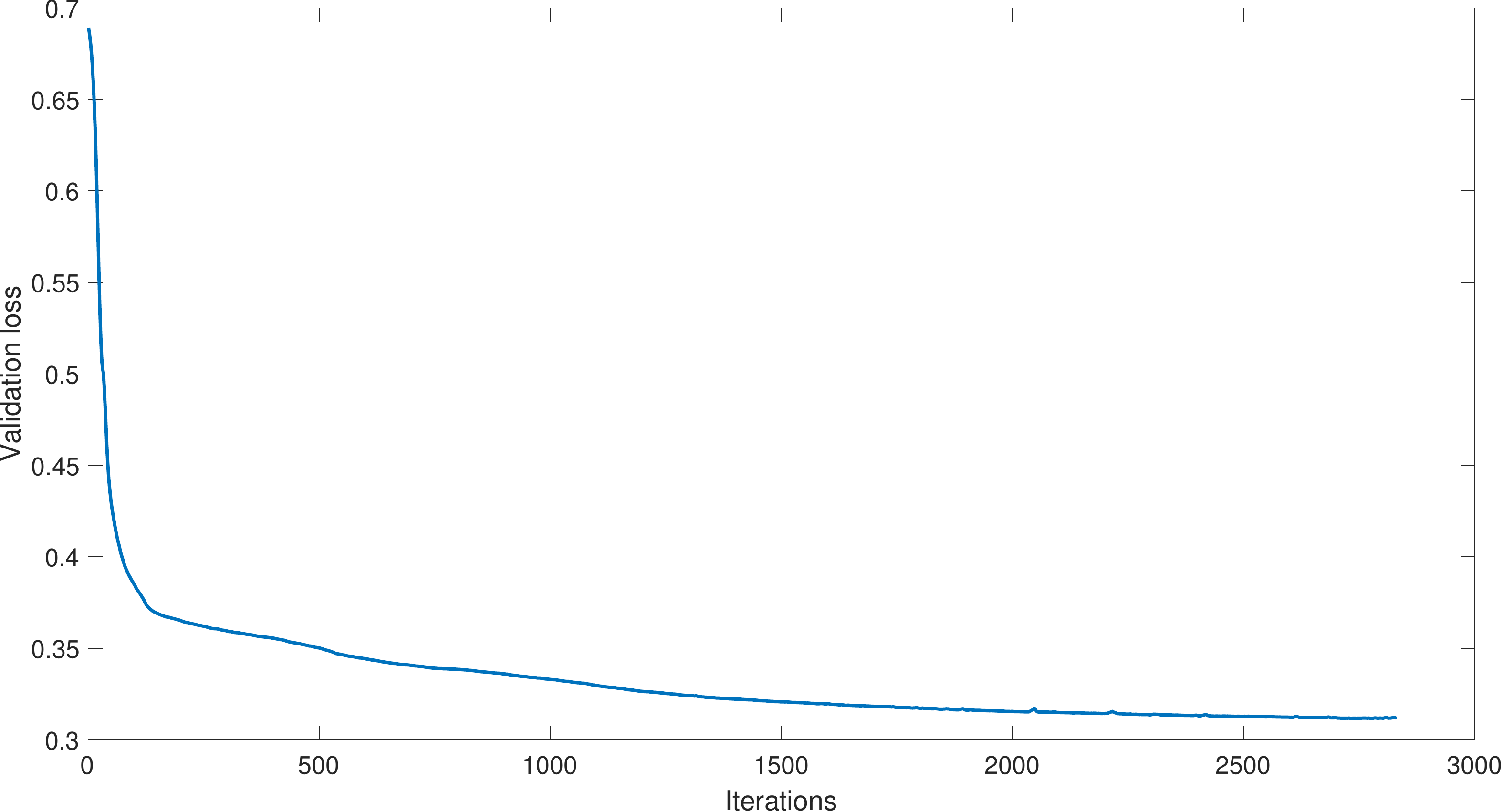}
\caption{Example \ref{exa:Bayesian deep neural net}: The NAGVAC method in Algorithm \ref{alg:ChapterFFVB:GVB-factor} for deep neural net modelling. 
The plot shows the validation loss over iterations.}
\label{fig:ChapterFFVB_NAGVAG}
\end{figure}

\end{example}

%-------------------------------------------------------------%
\subsection{Practical recommendation}
%-------------------------------------------------------------%
We conclude this section with a few practical recommendations that have been found useful in practice.
First, the fixed learning rate $\eps_0$ in \eqref{eq:scalar learning rate} requires some effort to tune, often based on trial and error.
Good starting values are $\eps_0=0.01$ or $\eps_0=0.001$, then adjusted after a few runs.
If the VB algorithm converges too quickly, it is probably because $\eps_0$ is set too large and needs to be reduced.  
Plotting the moving averaged lower bounds is a convenient and useful way for implementation diagnostic.
If this plot fluctuates too much, then a larger number of samples $S$ is needed, and also a wider moving average window $t_W$ should be used.
If these moving averaged lower bounds show a clear trend of decreasing, then something must have gone wrong.

It is often useful to standardize the data before model fitting.
For example, in regression modelling, each numerical column in the input matrix $X$ should be standardized to have mean zero and standard deviation of 1.

For challenging applications, it is a good idea to run the FFVB algorithm several times with different initialization $\lambda^{(0)}$ and select the one that ends up with the largest final lower bound.
It is also common to fix the random seed so that the results are reproducible. 

Finally, a simple practice known as {\it gradient clipping}\index{gradient clipping} is often found useful.
Gradient clipping makes the gradient estimate more well behaved by clipping its length while still maintaining its direction. 
It replaces the lower bound gradient estimate $\wh{\nabla_\l\LB}(\l)$ by 
\beq\label{eq:gradient clipping}
\frac{\ell_\text{threshold}}{\|\wh{\nabla_\l\LB}(\l)\|}\wh{\nabla_\l\LB}(\l),
\eeq
if the $\ell_2$-norm $\|\wh{\nabla_\l\LB}(\l)\|$ is larger than some threshold $\ell_\text{threshold}$, such as 100 or 1000.
Note that we used gradient clipping in Examples \ref{exa:Example 4} and \ref{exa:Bayesian deep neural net}. 

%===========================================================%
\subsection{A quick note on the bibliography of FFVB}\label{sec:bibliography}
%===========================================================%
This isn't a review paper, we therefore made no attempt to give a comprehensive literature review on Variational Bayes.
In addition to Section \ref{sec:MFVB:remarks}, this section is to give a short list of further reading on FFVB for the interested reader.
Compared to MFVB, FFVB is developed more recently with great contributions not only from machine learning but also the statistics community.
Further reading on control variate can be found in \cite{Paisley:2012,Nott:2012,Ranganath:2014,Tran:2017}.
See \cite{kingma2014adam,duchi2011adaptive} and \cite{zeiler2012adadelta} for the adaptive learning methods.
The natural gradient is first introduced in statistics, in the context of MLE, by \cite{Rao:1945}, popularized in machine learning by \cite{Amari:1998},
and developed further for applications in Variational Bayes by \cite{Sato:2001,Hoffman:2013,Martens:2014,Khan.Lin:2017,Lin:2019} and \cite{Tran:2020}.
The reparameterization trick can be found in \cite{Kingma:2014,Titsias:2014}.
The Cholesky GVB in Section \ref{sec: GVB Cholesky} is borrowed from \cite{Titsias:2014} and \cite{Tan:2018},
and more details of Algorithm \ref{alg:GVB natural gradient} together with the deep learning model in Example \ref{exa:Bayesian deep neural net} can be found in \cite{Tran:2020}.

There are more advanced variants of FFVB, such as the Importance Weighted Lower Bound of \cite{Burda.et.all:2016} and manifold VB of \cite{Tran.Nguyen.Nguyen:2020}, that aren't presented in this tutorial.
Also, there are recent advances in theoretical properties of VB approximations that we don't cover here; the interested reader is referred to \cite{Alquier.Ridgway:2019} and \cite{ZhangGao:2019}.

%===========================================================%
\section{VBLab software package and its applications}\label{sec:software package}
%===========================================================%
This section describes our end-user software package, VBLab, that implements several general FFVB algorithms described in Section \ref{sec:FFVB},
and demonstrates their use.  
The package also implements several other FFVB algorithms, such as the VAFC of \cite{Ong:2018} and manifold VB of \cite{Tran.Nguyen.Nguyen:2020}, that are not described in Section \ref{sec:FFVB}. 

VBLab is a probabilistic programming software package, currently available in Matlab, allowing automatic variational Bayesian inference on 
many pre-defined common statistical models and also user-defined models. The package provides various FFVB methods and works efficiently for high dimensional and complex posterior distributions. Users are not required to know the technicality behind the VB techniques provided; all they need to do is to supply their statistical model, which can be specified flexibly in various ways. 

%-------------------------------------------------------------%
\subsection{Bayesian logistic regression}
\label{sec:Bayesian logistic regression}
%-------------------------------------------------------------%
We consider again the Bayesian logistic regression model in Example \ref{exa:Example 4} and demonstrate how to use the VBLab package to output a VB approximation of the posterior distribution using the Cholesky GVB.

First, the Labour Force Participation dataset is loaded by calling the \lstinline{readData()} function with \lstinline{'LabourForce'} string as its input argument:
\begin{lstlisting}
% Load the Labour Force Participation dataset
labour = readData('LabourForce',...
                  'Intercept',true);  % Add column of 1 as intercept 
\end{lstlisting}
This dataset, together to several others, are included in the package and can be loaded using the \lstinline{readData()} function of the VBLab package.
For the purpose of this example, we use the entire Labour Force Participation data to train the model; if necessary, users can split the data into a training and testing data using the \lstinline{trainTestSplit()} function. 

Next, we create a logistic regression model object which is an instance of the \lstinline{LogisticRegression} class as follows:
\begin{lstlisting}
% Number of parameters of the Logistic Regression model
n_features = size(labour,2)-1;

% Create a Logistic Regression model object
Mdl = LogisticRegression(n_features,...
					     'Prior',{'Normal',[0,50]});
\end{lstlisting}
The \lstinline{LogisticRegression} model class requires at least one input argument indicating the number of model parameters.   
%The optional argument \lstinline{'Intercept'}, set to be \lstinline{true} by default, indicates whether or not a column of 1s will be added to the design matrix $X$; i.e. if $X$ already includes this column then \lstinline{'Intercept'} must be set to \lstinline{false}. 
The optional argument \lstinline{'Prior'} sets the prior for each coefficient of the regression model; here, a normal prior with zero mean and variance $50$ is used. By default, \lstinline{'Prior'} is set to be the standard normal distribution. The VBLab package provides commonly-used prior distributions including Normal, Uniform, Beta, Exponential, Gamma, Inverse-Gamma, Binomial and many others.

Given the logistic regression model object \lstinline{Mdl}, we now can call any FFVB algorithm provided in the VBLab package to produce a variational approximation of the posterior distribution. The following code calls the Cholesky GVB algorithm class \lstinline{CGVB}: 
\begin{lstlisting}
% Run Cholesky GVB 					
Post_CGVB = CGVB(Mdl,labour,...
			     'LearningRate',0.002,...  % Learning rate
			     'NumSample',50,...        % Number of VB samples
			     'MaxPatience',20,...      % For Early stopping
			     'MaxIter',5000,...        % Maximum number of iterations
			     'InitMethod','Custom',... % Randomly initialize variational mean   
			     'GradWeight1',0.9,...     % Momentum weight 1
			     'GradWeight2',0.9,...     % Momentum weight 2
			     'WindowSize',50,...       % Smoothing window for lowerbound      
			     'GradientMax',10,...      % For gradient clipping     
			     'LBPlot',true);           % Plot the lowerbound when finish
\end{lstlisting} 
The algorithm class \lstinline{CGVB} requires several input arguments specifying how this VB algorithm is implemented. The first argument is the statistical model of interest \lstinline{Mdl}, which can be defined as a class object or a function handle. The second argument is the dataset \lstinline{labour}, which can be either a Matlab table, or a single matrix with the last column to be the response data $y$.
%I comment this out as this sentence reads badly. I don't understand it at all.   
%As the \lstinline{CGVB} class, and also other VB classes of the VBLab package, will transfer this input \lstinline{data} to provided model object's methods or function handles calculating model related components such as log-likelihood or log-prior, users only need to make sure that the same data structure should used in the custom class's methods or function handles\footnote
Table \ref{tab:arguments} lists all the optional arguments of the \lstinline{CGVB} class  together with their equivalent mathematical notations and default values. 
%The \lstinline{'MaxIter'} argument specifies the maximum number of iterations of updating variational parameters. The \lstinline{'GradientMax'} argument is to set upper bound for the gradient of the lower bound to prevent the exploding gradient. The \lstinline{'InitMethod'} argument determines how the VB algorithm initialize the variational parameters. In this example, we set \lstinline{'InitMethod'} to be \lstinline{'MLE'}, implying that we use the maximum likelihood estimates of the model parameters as the initial values of the variational mean. The \lstinline{'LBPlot'} argument is set to be \lstinline{true}, specifying that we want to plot the smoothed lower bound at the end of the algorithm. The plot of the lower bound is useful to access the performance and correctness of the VB algorithm.     
\begin{table}[h]
	\begin{center}
		\begin{tabular}{cccl}
			\hline\hline
			\rule{0pt}{3ex}
			\bol{Argument}       &\bol{Default value}  & \bol{Notation}  & \bol{Description} \\
			\hline
			\rule{0pt}{3ex}
			\texttt{LearningRate}  & 0.002        & $\epsilon_0$   & Fixed learning rate in \eqref{eq:scalar learning rate}\\
			\texttt{NumSample}     & 50           & $S$          & Monte Carlo samples\\
			\texttt{MaxPatience}   & 20           & $P$          & Maximum patience\\
			\texttt{GradWeight1}   & 0.9          & $\beta_1$   & Adaptive learning weight \\
			\texttt{GradWeight2}   & 0.9          & $\beta_2$   & Adaptive learning weight \\
			\texttt{WindowSize}    & 50           & $t_W$        & Rolling window size\\
			\texttt{StepAdaptive}  & \texttt{MaxIter}/2 & $\tau$  & Threshold to start reducing learning rates \\
			\texttt{MaxIter}       & 1000         &              & Maximum number of iterations\\
			\texttt{GradientMax}   & 10           &     $\ell_\text{threshold}$         & Gradient clipping threshold in \eqref{eq:gradient clipping}\\ 
			\texttt{InitMethod}    & Random       &              & Initialization method\\
			\texttt{LBPlot}        & true         &              & Whether or not to plot the lower bounds\\
			\hline\hline
		\end{tabular}
	\end{center}
	\caption{Input arguments of the \lstinline{CGVB} class constructor with their equivalent mathematical notations and default values.}
	\label{tab:arguments}
\end{table}

\begin{table}[h]
	\begin{center}
		\begin{tabular}{clc}
			\hline\hline
			\rule{0pt}{3ex}
			\bol{Output}      & \bol{Description}   & Notation\\
			\hline
			\rule{0pt}{3ex}
			\texttt{LB}            & The lower bound estimated in each iteration   & $\widehat{\text{LB}}(\lambda)$\\
			\texttt{LB\_smooth}    & The smoothed lower bound estimated in each iteration        &  $\overline{\text{LB}}(\lambda)$ \\
			\texttt{mu}            & Mean of the Gaussian variational distribution        & $\mu$ \\
			\texttt{L}             & The lower triangular matrix of the variational covariance matrix  &  $L$\\
			\texttt{Sigma}         & The variational covariance matrix   & $\Sigma$\\
			\texttt{sigma2}        & The diagonal of the variational covariance matrix    &  $\text{diag}(\Sigma)$  \\
			\hline\hline
		\end{tabular}
	\end{center}
	\caption{Outputs of the CGVB algorithm together with their description and equivalent mathematical notations. }
	\label{tab:cgvb_output}
\end{table}

The outputs of the CGVB algorithm class are store in the attribute \lstinline{Post}, which is a Matlab structure data type, of the output \lstinline{Post_CGVB}. Table \ref{tab:cgvb_output} lists the fields of the \lstinline{Post} attribute together with their descriptions and equivalent notations in Section \ref{sec: GVB Cholesky}. 
For example, the following code shows how to extract the mean $\mu$ and variance $\text{diag}(\Sigma)$ of 
the Gaussian variational distribution,
and then plots the corresponding normal density using \lstinline{vbayesPlot()} function of the VBLab package, as shown in Figure \ref{f:example_4}:
\begin{lstlisting}
% Extract variational mean and variance
mu_vb = Post_CGVB.Post.mu;           % Varational mean
sigma2_vb = Post_CGVB.Post.sigma2;   % Variational variance

% Plot the variational distribution of each model parameter
for i=1:num_feature
    subplot(3,3,i)
	vbayesPlot('Density',...
			   'Distribution',{'Normal',[mu_vb(i),sigma2_vb(i)]})
end
\end{lstlisting}
We can also extract the smoothed lower bounds from \lstinline{Post} and plot them as shown in the last panel of Figure \ref{f:example_4}:
\begin{lstlisting}
% Plot the smoothed lower bound
subplot(3,3,9)
plot(Post_CGVB.Post.LB_smooth)
title('Lower bound')
\end{lstlisting}

\begin{figure}[h]
	\centering
	\includegraphics[width=1\columnwidth]{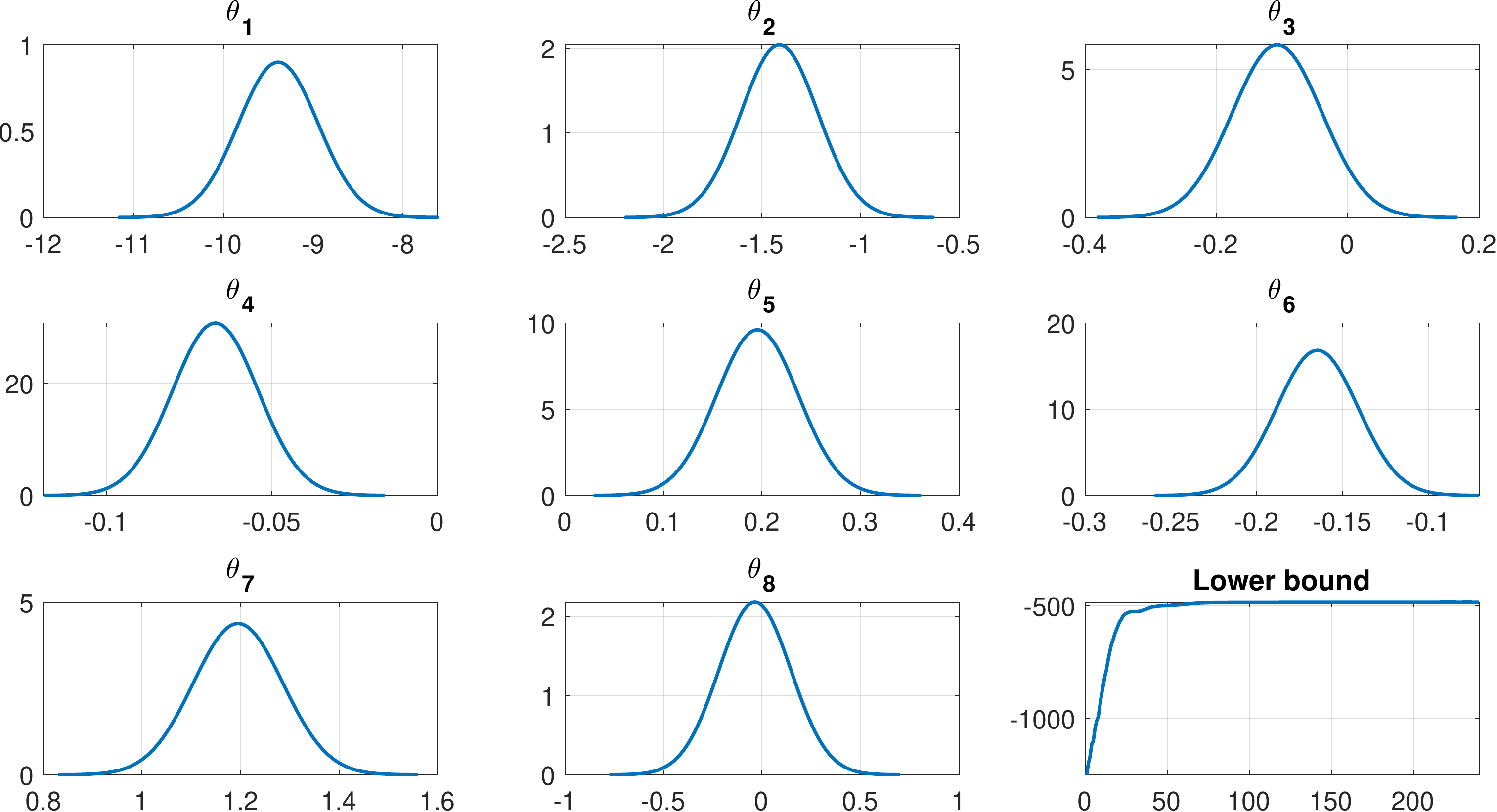}
	\caption{Variational distribution densities of model parameters and the smoothed lower bound.}
	\label{f:example_4}
\end{figure}

%-------------------------------------------------------------%
\subsection{Bayesian deep neural networks}
%-------------------------------------------------------------%
This section demonstrates how to use the VBLab package for variational Bayesian inference in Bayesian deep neural networks.
The package implements the DeepGLM model of \cite{Tran:2020}, which provides a unified framework for flexible regression that combines the deep neural network method in machine learning for data representation 
with the popular Generalized Linear Models (GLM) in statistics. This section also describes the use of the \lstinline{VAFC} class that implements the GVB with factor covariance (VAFC) algorithm briefly mentioned in Section \ref{sec: GVB factor}. 

We first load the German Credit data by calling the \lstinline{readData()} function with \lstinline{'GermanCredit'} string as its input argument:
\begin{lstlisting}
% Load the German Credit dataset
credit = readData('GermanCredit',...
			      'Type','Table',...    % Store data in a table
			      'Intercept',true,...  % Add column of 1 for intercept
			      'Normalized',true);   % Neural Networks work more efficient 
			                            % with normalized data

% Number of input features
n_features = size(credit,2) - 1;
\end{lstlisting}

We then define an instance of the \lstinline{DeepGLM} model class, which specifies important components such as the prior, likelihood function, etc., for the DeepGLM model:
\begin{lstlisting}
% Define a deepGLM model object
Mdl = DeepGLM([n_features,10,10],...
			  'Activation','Relu',...
			  'Distribution','Binomial');
\end{lstlisting}
The \lstinline{DeepGLM} model class requires at least one input argument which is the structure of the neural network. 
The code above specifies a structure that has one input layer with \lstinline{n_features} units (including the bias term), and two hidden layers each with 10 units.
The \lstinline{'Activation'} argument, set to \lstinline{'Relu'} by default, specifies the activation function used for each hidden unit. 
The \lstinline{'Distribution'} argument, is \lstinline{'Normal'} by default, specifies the distribution used for the response data.
In the code above, we set \lstinline{'Distribution'} to be \lstinline{'Binomial'} as the response variable in the German Credit data is binary.
Users are referred to the documentation of the VBLab package for a more comprehensive discussion on the \lstinline{DeepGLM} class.   

Finally, we run the \lstinline{VAFC} algorithm class to approximate the posterior distribution of this DeepGLM model (for bigger DeepGLM models or if computational speed-up is of primary importance, one should use the \lstinline{NAGVAC} algorithm class rather than \lstinline{VAFC}):
\begin{lstlisting}
% Run VAFC to obtain VB approximation of the posterior
Post_VAFC = VAFC(Mdl,credit,...
				 'Validation',0.2,...
				 'LearningRate',0.002,... 
				 'NumFactor',4,...
				 'NumSample',50,...
				 'GradWeight',0.9,...
				 'MaxPatience',100,...
				 'MaxIter',10000,...
				 'GradientMax',200,...
				 'WindowSize',30,...
				 'InitMethod','Random');
\end{lstlisting}
The \lstinline{'NumFactor'} argument, 4 in this example, specifies the number of factors used in VAFC.
As we use a prediction loss on a validation dataset to assess the convergence of the VAFC algorithm, we split \lstinline{data} into a training set, for parameter estimation, and a validation set, for early stopping. The \lstinline{'Validation'} argument, set to be $0.2$ in this example, indicates that we use $20\%$ of the data to form the validation set. 

%-------------------------------------------------------------%
\subsection{Volatility modelling with the RECH models}
%-------------------------------------------------------------%
Let $y=\{y_t,\ t=1,...,T\}$ be a time series of financial asset returns and $\mathcal F_t$ be the $\sigma$-field of the information up to time $t$. 
Volatility, defined as the conditional variance $\sigma_t^2:=\Var(y_t|\mathcal F_{t-1})$, is of high interest in the financial sector.
Conditional heteroskedastic models, such as GARCH of \cite{Bollerslev:1986}, represent $\sigma_t^2$ as a deterministic function of the observations and conditional variances in the previous time steps. 
\cite{Nguyen:2020} recently propose a new class of conditional heteroskedastic models, namely the REcurrent Conditional Heteroskedastic (RECH) models, by combining recurrent neural networks (RNNs) and GARCH-type models, for flexible modelling of the volatility dynamics.
The conditional variance in the RECH models is the sum of two components: the recurrent component modeled by an RNN, and the garch component modeled by a GARCH-type structure. 
For example, by using the Simple Recurrent Network (SRN) for the recurrent component $\omega_t$ and the standard GARCH(1,1) for the garch component, they obtain the SRN-GARCH specification of the RECH models as:
\begin{subequations}
	\begin{align}
	y_t &= \sigma_t\eps_t,\;\;\eps_t\stackrel{iid}{\sim}\N(0,1),\;\;t=1,2,...,T  \label{RNN-GARCH1}\\
	\sigma_t^2 &= \omega_t+ \alpha y_{t-1}^2 + \beta \sigma_{t-1}^2,\;\;t=2,...,T,\;\; \sigma^2_1=\sigma^2_0             \label{RNN-GARCH2}\\
	\omega_t&=\beta_0+\beta_1 h_t,\;\; t=2,...,T,\label{RNN-GARCH3}\\
	h_t&=\phi(v x_t  + wh_{t-1}+b),\;\;t=2,...,T,\;\; \text{with} \;\; h_1 \equiv 0;\label{RNN-GARCH4}
	\end{align}
\end{subequations}
\cite{Nguyen:2020} suggest $x_t=(\omega_{t-1},y_{t-1},\sigma^2_{t-1})^\top$. 
The SRN-GARCH model has 7 parameters: $\theta = (\alpha,\beta,\beta_0,\beta_1,v,w,b)$. 

The following code shows how to use the VBLab package for Bayesian inference in RECH using the Manifold GVB method of \cite{Tran.Nguyen.Nguyen:2020}. First, we read the SP500 data by calling the \lstinline{readData()} function
\begin{lstlisting}
% Load the SP500 daily return data
sp500 = readData('RealizedLibrary',...
				 'Index','SP500',...  
				 'Length',1000);        % Extract only the last 1000 observations
\end{lstlisting}
In this example, we use only the last $1000$ observations to perform the approximation Bayesian inference by setting the value of the \lstinline{'Length'} argument to be  $1000$.

Next, we define a RECH model together with its prior:
\begin{lstlisting}
% Define priors for model parameters using 2D cell array
% Parameter names must be specified correctly
prior = {{'v','w','b'},'Normal',[0,1];...
		 {'beta0','beta1'},'Inverse-Gamma',[0.25,2.5];...
		 {'alpha','beta'},'Uniform',[0,1]};

% Define a RECH model with SRN-GARCH specification		 
Mdl = RECH('SRN-GARCH',...
		   'Prior', prior);
\end{lstlisting}
We define the priors for the SRN-GARCH's parameters using a Matlab 2D cell array. Each row of this cell array has three elements including: parameter names, name of the prior distribution and its parameters. The parameter names are put in a 1D cell array listing the model parameters that share the same prior distribution. The prior distribution name must be one of the  distribution classes available in the VBLab package. The distribution parameters must be stored in a Matlab 1D array. In this example, we use the same priors as suggested in \cite{Nguyen:2020}. The model class \lstinline{RECH} requires at least one input argument, which is a particular specification of the RECH models. The current version of the VBLab package provides three specifications for the RECH models including \lstinline{'SRN-GARCH', 'SRN-GRJ'} and \lstinline{'SRN-EGARCH'}. The \lstinline{'Prior'} argument sets the priors for model parameters defined previously in the variable \lstinline{prior}. Users can refer to the documentation of the VBLab package for more comprehensive discussion on the \lstinline{RECH} model class.

Given the model object \lstinline{rech_model} defined by the \lstinline{RECH} model class, we now run the Manifold GVB method by calling the \lstinline{MGVB} algorithm class:
\begin{lstlisting}
% Run MGVB given the data and RECH model 
Post_MGVB = MGVB(Mdl,y,...
				 'NumSample',100,...
				 'LearningRate',0.01,...
				 'GradWeight',0.4,...
				 'MaxPatience',50,...
				 'MaxIter',2500,...
				 'GradientMax',100,...
				 'WindowSize',30);
\end{lstlisting}

Similar to the other VB algorithm classes, the \lstinline{MGVB} class stores the outputs in a Matlab structure which can be used as shown in the following code to visualize the density of variational distribution and smoothed lower bound.
\begin{lstlisting}
% Extract variation mean and variance
mu_vb     = Post_MGVB.Post.mu;
sigma2_vb = Post_MGVB.Post.sigma2;

% Define parameter names for plotting
param_name = {'\beta_0','\beta_1','\alpha','\beta','v','w','b'};

% Plot the variational distribution of each parameter
for i=1:num_feature
	subplot(3,3,i)
	vbayesPlot('Density',...
			   'Distribution',{'Normal',[mu_vb(i),sigma2_vb(i)]})
	title(param_name{i})
end

% Plot the smoothed lower bound
subplot(3,3,9)
plot(Post_MGVB.Post.LB_smooth)
title('Lower bound')
\end{lstlisting}

\begin{figure}[h]
	\centering
	\includegraphics[width=1\columnwidth]{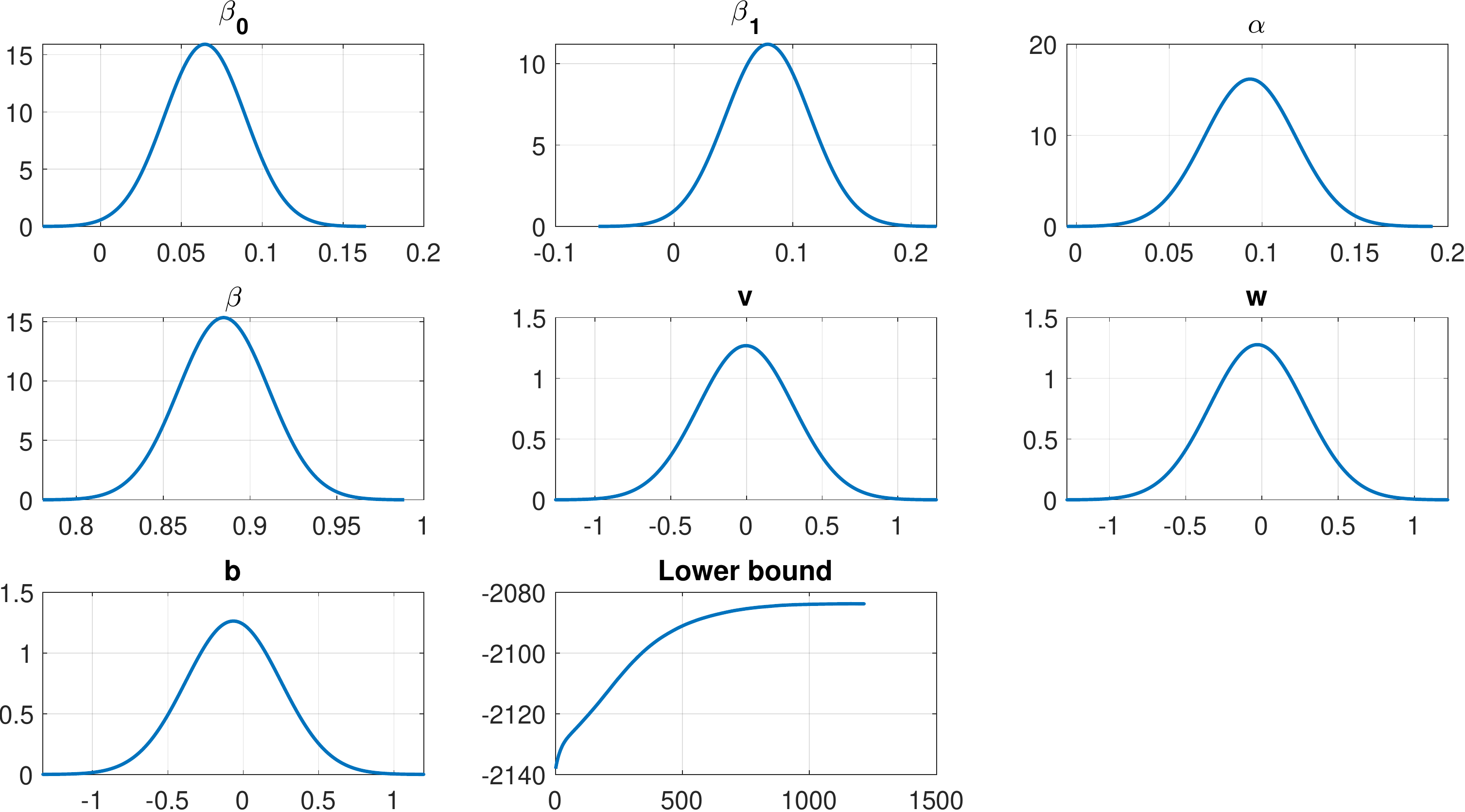}
	\caption{Variational distribution densities of the RECH model parameters and the smoothed lower bound.}
	\label{f:example_rech}
\end{figure}

%-------------------------------------------------------------%
\subsection{Using the VBLab package for user-defined models}
%-------------------------------------------------------------%   
For pre-defined models such as logistic regression or DeepGLM, we can use the model classes provided 
in the VBLab package to create the corresponding model object before calling a VB algorithm as demonstrated in the previous sections.
For user-defined statistical models, the package provides several ways for custom-built models that work with VB algorithm classes such as \lstinline{CGVB}, \lstinline{VAFC}, \lstinline{MGVB} and \lstinline{NAGVAC}. 
It only requires users to specify a function to compute $h(\theta)$ and $\nabla_\theta h(\theta)$ as discussed in Algorithm \ref{alg:GVB-Cholesky} and \ref{alg:ChapterFFVB:GVB-factor}. 
We demonstrate this use of the package below using logistic regression.

\subsubsection{Bayesian logistic regression with manual gradient}
Users need to supply a function that evaluates the model-specific term $h(\theta)$.
For VB algorithms that are based on the reparameterization trick, users also need to supply the gradient $\nabla_\theta h(\theta)$.
This section considers the case where $\nabla_\theta h(\theta)$ can be calculated manually,
and Section \ref{sec:AD} demonstrates how to use Automatic Differentiation to calculate this gradient.

The following code defines a function that computes both $h(\theta)$ and $\nabla_\theta h(\theta)$ as in \eqref{eqn:h_function_ex4}-\eqref{eqn:grad_h_function_ex4_1}:
\begin{lstlisting}
function [h_func_grad,h_func] = grad_h_func_logistic(data,theta,mdl)

    % Extract additional settings
	d = length(theta);
	sigma2 = mdl.Prior(2);
	
	% Extract data
    X = data(:,1:end-1);
	y = data(:,end));
	
	% Compute log likelihood
	aux = X*theta;
	llh = y.*aux-log(1+exp(aux));
	llh = sum(llh);  
	
	% Compute gradient of log likelihood
	ppi       = 1./(1+exp(-aux));
	llh_grad  = X'*(y-ppi);
	
	% Compute log prior
	log_prior =-d/2*log(2*pi)-d/2*log(sigma2)-theta'*theta/sigma2/2;
	
	% Compute gradient of log prior
	log_prior_grad = -theta/sigma2;
	
	% Compute h(theta) = log p(y|theta) + log p(theta)
	h_func = llh + log_prior;
	
	% Compute gradient of the h(theta)
	h_func_grad = llh_grad + log_prior_grad;
	
	% h_func_grad must be a column
	h_func_grad = reshape(h_func_grad,length(h_func_grad),1);

end 
\end{lstlisting}

There are some rules to define a proper function for calculating $h(\theta)$ and $\nabla h(\theta)$ that it is compatible with the VB algorithm classes in the package.
\begin{itemize}
	\item The input should have three arguments:
	\begin{itemize}
		\item \lstinline{data}: The data that is used for calculating $h(\theta)$ and $\nabla_\theta h(\theta)$. 
		\item \lstinline{theta}: A \textit{column} vector of model parameters.  
		\item \lstinline{mdl}: Any additional setting necessary for defining the custom-built models. This \lstinline{mdl} variable must be created before running VB algorithms and used as the input to the \lstinline{'Setting'} argument of the VB algorithm classes.
	\end{itemize}
	\item There are two outputs:
	\begin{itemize}
		\item The first output, e.g. \lstinline{h_func_grad} as in the previous code, must be a \textit{column} vector that returns the value of  
		$\nabla_\theta h(\theta)$.
		\item The second output, e.g. \lstinline{h_func} as in the previous code, must be a \textit{scalar} that returns the value $h(\theta)$.
	\end{itemize}
\end{itemize} 
To assist with calculating gradient, VBLab provides the static methods \lstinline{logPdfFnc()} and \lstinline{GradlogPdfFnc()}
for conveniently computing the log density and its gradient for common prior distributions. 
For example, rather than having to specify log normal density and its gradient explicitly as in the code above, 
one can call the functions \lstinline{Normal.logPdfFnc(theta,mu,sigma2)} and \lstinline{Normal.GradlogPdfFnc(theta,mu,sigma2)} to compute the log density and its gradient, respectively, of the Gaussian distribution with mean \lstinline{mu} and variance \lstinline{sigma2}.  

Given the function to compute $h(\theta)$ and $\nabla h(\theta)$, we now use a VB algorithm class, e.g. \lstinline{CGVB}, to produce a variational approximation of the posterior distribution defined by  $h(\theta)$:
\begin{lstlisting}
% Load the Labour Force Participation dataset
labour = readData('LabourForce',...
				  'Type','Matrix',...
				  'Intercept',true);

% Number of model parameters. Adding 1 for the intercept. 
n_features = size(labour,2)-1;

% Struct to store prior
setting.Prior = [0,50];

% Initialize the variational mean
mu_init = normrnd(0,0.01,n_features,1);

% Create an CGVB object and run the CGVB algorirthm
Post_CGVB = CGVB(@grad_h_func_logistic,labour,...
				 'NumParams',n_features,...    
				 'Setting',setting,...
				 'MeanInit',mu_init,...
				 'LearningRate',0.002,...  
				 'NumSample',50,...        
				 'MaxPatience',20,...      
				 'MaxIter',5000,...          
				 'GradWeight1',0.9,...    
				 'GradWeight2',0.9,...    
				 'WindowSize',50,...       
				 'GradientMax',10,...           
				 'LBPlot',true);     
\end{lstlisting}

After loading the data, we create the structure \lstinline{setting} to store additional variables, rather than the data and model parameters, necessary for computing $h(\theta)$ and $\nabla_\theta h(\theta)$. 
The handle of the user-defined function \lstinline{grad_h_func_logistic} is passed to the \lstinline{CGVB} class constructor as the first input argument.  
We also need to set the value of the argument \lstinline{'NumParams'} to be the number of model parameters and pass the variable \lstinline{setting} to the \lstinline{'Setting'} argument. 
The \lstinline{CGVB} class provides several ways to initialize the variational mean $\mu$. In this example, we initialize $\mu$ randomly using a normal distribution, which is used as the input to the \lstinline{'MeanInit'} argument. 
The other algorithmic arguments of the \lstinline{CGVB} class are set as in Section \ref{sec:Bayesian logistic regression}.

\subsubsection{Bayesian logistic regression with Automatic Differentiation}\label{sec:AD}
Instead of computing the gradient $\nabla_\theta h(\theta)$ manually as in the previous section,
we can compute it using Matlab's Automatic Differentiation facility,
which is a technique for evaluating derivatives numerically and automatically.
The general rule of using Automatic Differentiation in Matlab is that we must call \lstinline{dlgradient()} inside a helper function,
and then evaluate the gradient using \lstinline{dlfeval()}.
The following code modifies the function \lstinline{grad_h_func_logistic()} above to output $\nabla_\theta h(\theta)$ using Automatic Differentiation.
\begin{lstlisting}
% Define a function to compute h(theta) = log p(theta) + log p(y|theta)
function h_func = h_func_logistic(data,theta,mdl)

	% Extract additional settings
	d = length(theta);
	sigma2 = mdl.Prior(2);
	
	% Extract data
	X = data(:,1:end-1);
	y = data(:,end));
	
	% Compute log likelihood
	aux = X*theta;
	log_lik = y.*aux-log(1+exp(aux));
	log_lik = sum(log_lik);  
	
	% Compute log prior
	log_prior =-d/2*log(2*pi)-d/2*log(sigma2)-theta'*theta/sigma2/2;
	
	% h = log p(y|theta) + log p(theta)
	h_func = llh + log_prior;

end
\end{lstlisting}

\begin{lstlisting}
% Define a function to call dlgradient to automatically compute the gradient
% of the h function
function [h_func_grad,h_func] = grad_h_func_logistic_AD(data,theta,mdl)

	h_func = h_func_logistic(data,theta,mdl);
	h_func_grad = dlgradient(h_func,theta);    

end
\end{lstlisting}

\begin{lstlisting}
function [h_func_grad,h_func] = grad_h_func_logistic(data,theta,mdl)

    % Convert parameters to dlarray data type
	theta_AD = dlarray(theta);
	
	% Evaluate the function containing dlgradient using dlfeval
	[h_func_grad_AD,h_func_AD] = dlfeval(@grad_h_func_logistic_AD,data,theta_AD,mdl);
	
	% Convert parameters from dlarray to matlab array
	h_func_grad = extractdata(h_func_grad_AD);
	h_func = extractdata(h_func_AD);
	
	% Make sure the output is a column vector
	h_func_grad = reshape(h_func_grad,length(h_func_grad),1);

end 
\end{lstlisting}
This \lstinline{grad_h_func_logistic} now can be used as the first input argument of the \lstinline{CGVB} algorithm class as before.

%The \lstinline{grad_log_lik_logistic()} is the function in which \lstinline{dlgradient()} is called to automatically compute the gradient of $\text{log} \; p(y|\theta)$, which is the output of the \lstinline{log_lik_logistic()} function. Inside the \lstinline{grad_h_func_logistic()} function, we run AD to compute $\Delta \text{log} \; p(y|\theta)$ by evaluating the \lstinline{grad_log_lik_logistic()} function, which contains \lstinline{dlgradient()}, using \lstinline{dlfeval()}. Before calling the \lstinline{dlfeval()} function, we have to convert the vector of model parameters \lstinline{theta} to the \lstinline{dlarray} data type using the \lstinline{dlarray()} function. After running AD with \lstinline{dlfeval()}, we have to convert the outputs to the Matlab array using the \lstinline{extractdata()} function.  

\bibliographystyle{apalike}
\bibliography{references_v1}

\end{document}